\documentclass[a4paper,12pt]{article}
\usepackage{amsmath}
\usepackage{amsthm}
\usepackage{amsfonts}
\usepackage{mathrsfs}
\usepackage{graphicx}
\usepackage{hyperref}
\usepackage{color}
\usepackage{tabularx}

\newcommand{\lR}{\mathrm{I\hspace{-0.7mm}R}}

\newtheorem{lemma}{Lemma}[section]
\newtheorem{theorem}[lemma]{Theorem}

\newtheorem{corollary}[lemma]{Corollary}

\numberwithin{equation}{section}

\hypersetup{ pdftitle={Draft},
pdfauthor={Fiki T. Akbar, Bobby E. Gunara}, 
pdfkeywords={Local Esistence, Dyonic Black Hole}, 
bookmarksnumbered, pdfstartview={FitH}, urlcolor=blue, }

\begin{document}
\pagestyle{plain}




\title{\LARGE\textbf{Static Spacetimes In Higher Dimensional Scalar-Torsion Theories With Non-Minimal Derivative Coupling}}

\author{ Bobby E. Gunara$^1$\footnote{Corresponding author},\and Mulyanto$^1$, \and Rahmat H. Alineng$^1$, \and Fiki T. Akbar$^1$, \and Hadi Susanto$^2$}
\date{$^1$Theoretical Physics Laboratory,\\ Theoretical High Energy Physics Research Division,\\ Faculty of Mathematics and Natural Sciences,\\  Institut Teknologi Bandung\\%
    $^2$Department of Mathematics, Khalifa University, PO Box 127788,\\ Abu Dhabi, United Arab Emirates\\[2ex]%
    \today}

\maketitle




\begin{abstract}

In this paper we consider a class of static spacetimes in higher dimensional ($D \ge 4$) scalar-torsion theories with non-minimal derivative coupling and the scalar potential turned on. The spacetime is conformal to a product space of a two-surface and a $(D-2)$-dimensional submanifold. Analyzing the equations of motion in the theory we find that the $(D-2)$-dimensional submanifold has to admit constant triplet structures in which the torsion scalar is one of them. This implies that these equations of motion can be simplified into a single highly non-linear ordinary differential equation called the master equation. Then, we show that in this case the solution admits at least a naked singularity at the origin which is not a black hole. In the asymptotic region, the spacetimes converge to spaces of constant scalar curvature which are generally not Einstein. We also use perturbative method to linearize the master equation and construct the first order solutions. At the end, we establish the analysis of local-global existences of the master equation and then, prove the non- existence of regular global solutions.

\end{abstract}

\newpage

\section{Introduction}	
\label{sec:intro}

Teleparallel gravity is an alternative theory of gravitational interaction where gravitation is attributed to torsion rather than curvature as in general relativity (GR). In this new theory, a vielbein and an alternative connection called  Weitzenb\"{o}ck connection serve as dynamical variables. The field equation of motions can be thought of as a force equation similar to   the Lorentz force equation in Maxwell electrodynamics which implies that there is no geodesic equations in the theory, see for example, \cite{Aldrovandi}.

Within this decade, this teleparallel theories  have been actively studied  due to several developments in the cosmological context. For example, general teleparallel formulation of gravity, that is $f(T)$ where $T$ is the torsion scalar, has been proposed as an alternative gravitational theory in which it gives several interesting cosmological models. In particular, we could have two models of the accelerated expansion of the late Universe, that is, without the presence of dark energy \cite{Bengochea:2008gz}, or with no need of fine tuning allowed by several cosmological observations \cite{Linder:2010py}. In the early Universe, we have an inflationary scenario without inflaton using the so-called Born-Infeld teleparallel formulation \cite{Ferraro:2006jd}. These models together with other possible models of $f(T)$ gravity have been reviewed in \cite{Cai:2015emx}. Keeping track of the GR limit is an interesting overall question in generalised theories of gravity. In \cite{Habib:2018}, they demonstrate that gravitational waves have the same polarization modes as General Relativity if the boundary term is minimally coupled to the torsion scalar and the scalar field.

Although this $f(T)$ gravity has gained some interesting results in cosmology, it is of interest to consider a static solution related to a black hole solution in the theory which is also a test  analogue to GR. Some authors show that in four dimensions the static spherically symmetric black holes do exist, see for example \cite{Wang:2011xf,Boehmer:2011gw, Ferraro:2011ks}. Another interesting feature of this theory is so called junction condition, that is, the matching condition of hypersurfaces of spacetimes, see for example \cite{delaCruz-Dombriz:2014zaa}. Recently, some authors have studied the existence of Birkhoff-like theorems in torsional theories, see for example \cite{delaCruz-Dombriz:2018vzn}. However, in higher dimensions we only have one example \cite{Capozziello:2012zj}. Particularly, our interest is to consider a minimal $f(T)$ gravity coupled to a scalar field in which we set $f(T) = T$. In \cite{kofinas2012torsi} the authors constructed a scalar-torsion theory where the torsion is coupled to a scalar with non-minimal derivative coupling and the scalar potential turned on. Then, they discuss the static spherically symmetric solutions of four dimensional scalar-torsion theories which can be simply described by so called master equation. However, as we read the paper, we found a crucial miscalculation which has been corrected in \cite{Yaqin2017comment} to produce the correct master equation. This implies that there is no wormhole-like solution as claimed in \cite{kofinas2012torsi} for a particular form of the scalar potential.

The purpose of this paper is to consider scalar-torsion theories in higher dimensions ($D \ge 4$) where $f(T) = T$ coupled non-minimally with the kinetic terms of a real scalar field $\phi$. In particular, the spacetime ${\mathcal M}^D$ is set to be static and conformal to ${\mathcal T}^2 \times {\mathcal S}^{D-2}$ where ${\mathcal T}^2 $ and $ {\mathcal S}^{D-2}$ are a two-surface and a $(D-2)$-dimensional submanifold, respectively. The scalar field $\phi$ depends only on the radial coordinate $r$. This setup has two consequences as the following. First, the $(D-2)$-dimensional submanifold ${\mathcal S}^{D-2}$ should admit constant triplet structures where the torsion scalar $\hat{T}$ belongs to them which further restricts ${\mathcal S}^{D-2}$. So far, some common examples do exist, such as 2-sphere $S^2$, the $(D-2)$-dimensional torus $T^{D-2}$, and $\lR^{D-2}$. The latter example has been considered in \cite{Capozziello:2012zj}. Second, all equations of motions in the theory can be simplified into a single non-linear ordinary differential equation called the master equation. This feature differs from the standard Einstein gravitational theory for static spacetimes. 
 
In this model we show that there exists a static solution with at least a naked singularity at the origin. Moreover, there could be more singularities besides the origin if  the mean curvature vanishes at another point which leads to the blow up of equations of motion and other geometrical quantities such as the Kretschmann and the Ricci scalars. In the asymptotic region where $x \to +\infty$, we assume the solution of the master equation $Y(x)$ to converge to a constant $Y_0$, which implies that in general the geometries become spaces of constant curvature which are not Einstein. In particular, the geometry converge to Einstein only for $D=4$ and $\epsilon =1$. 

As mentioned above, in the theory we have a master equation which is very difficult to solve. Therefore, we have to employ perturbative method in order to simplify the master equation into a linear equation. This method means that we expand the solution $Y(x) = Y_0 + Y_1(x)$ with $|Y_1| \ll |Y_0|$  in the asymptotic region. The function $Y_1$ is nothing but the solution of the linear version of the master equation which decreases exponentially  in the case of  $D \ge 4$.  In particular, for $D > 4$ this method restricts that we have only two consistent models. To make our analysis complete, we establish local-global existences of the master equation showing that a regular global solution does exist only for the case of $D \ge 6$.

We organize this paper as follows. In Section \ref{sec:scalartorsion}, we give a quick review of the scalar-torsion theories in higher dimensions including the introduction of some notations and the derivation of equations of motion. We discuss generally the static spacetimes and derive the master equation in Section \ref{sec:staticgenset}. We provide two simple solutions of the master equation in Section \ref{sec:specsolexact}. In section \ref{sec:propertiesstatic}, we discuss some properties of the solution including the non-existence of near-horizon limit, the asymptotic geometries, and the perturbative solution.  We construct the local-global existence analysis in Section \ref{sec:localglobalex}. Finally, we conclude our results in Section \ref{sec:conclusion}.

\section{Scalar-Torsion Theories in Higher Dimensions}
\label{sec:scalartorsion}
In this section we briefly review the scalar-torsion theory which can be viewed as an alternative gravitational theory called teleparallel formulation of gravity \cite{kofinas2012torsi}. In addition, we introduce some notations which are useful for our analysis in the paper.

\subsection{Short Review: Torsion and Curvature}
Our starting point is to define a quantity called torsion on a $D$-dimensional spacetime ${\mathcal M}^D$ whose form is given by 
\begin{equation}\label{torsion}
{T^{\lambda}}_{\mu\nu} = {\omega^{\lambda}}_{\nu\mu}-{\omega^{\lambda}}_{\mu\nu} ~ ,
\end{equation}  
where $ {\omega^{\lambda}}_{\mu\nu} $ is called alternative connection with curved indices $\mu, \nu, \lambda = 0, 1, ..., D$. We also introduce a contorsion which connects the alternative connection and the Christoffel symbol ${\Gamma^{\lambda}}_{\mu\nu} $ defined as
\begin{eqnarray}\label{contorsion}
\mathcal{K}_{\lambda\mu\nu}&=&\frac{1}{2}(T_{\nu\lambda\mu}-T_{\mu\nu\lambda}-T_{\lambda\mu\nu})\notag\\
&=&\omega_{\lambda\mu\nu} - \Gamma_{\lambda\mu\nu} ~ ,
\end{eqnarray}
where $\Gamma_{\lambda\mu\nu} = g_{\lambda\alpha} {\Gamma^{\alpha}}_{\mu\nu} $ and $g_{\lambda\alpha} $ is the spacetime metric endowed on ${\mathcal M}^D$. In this teleparallel formulation, one adds a tensor $ S^{\mu\nu\lambda} $ defined as
\begin{equation}\label{tensorS}
S^{\mu\nu\lambda}=\frac{1}{2}\mathcal{K}^{\nu\lambda\mu}+\frac{1}{2}(g^{\mu\lambda}{T_{\rho}}^{\rho\nu}-g^{\mu\nu}{T_{\rho}}^{\rho\lambda})=-S^{\mu\lambda\nu} ~ ,
\end{equation}
such that we can define the torsion scalar 
\begin{eqnarray}\label{torsionscalar}
T&=&S^{\mu\nu\lambda}T_{\mu\nu\lambda}\notag\\
&=&\frac{1}{4}T^{\mu\nu\lambda}T_{\mu\nu\lambda}+\frac{1}{2}T^{\mu\nu\lambda}T_{\lambda\nu\mu}-{T_\nu}^{\nu\mu}{T^{\lambda}}_\lambda\mu ~ .
\end{eqnarray}
Similar to general relativity, the connection $ {\omega^{\lambda}}_{\mu\nu} $ can be written in term of vielbein $ e_a $ and its dual $ e^a $, namely,
\begin{equation}\label{altconnviel}
\omega^\lambda_{~ \mu\nu}={e_{a}}^\lambda{e^a}_\mu,_\nu ~ ,
\end{equation}
referred to as Weitzenb\"ock connection which implies
\begin{equation}\label{torsionviel}
{T^\lambda}_{\mu\nu}=-{e_{a}}^\lambda({e^a}_\nu,_\mu-{e^a}_\mu,_\nu) ~ .
\end{equation}
where $ e_a $ and $ e^a $ satisfy
\begin{equation}\label{identviel}
{e^a}_\mu{e^{\nu}}_a=\delta^\nu_\mu ~,\qquad {e^a}_\mu{e^{\mu}}_b=\delta^a_b ~ ,
\end{equation}
with $a, b, c = 0, 1,....,D$ are the flat indices. Here, we have the metric tensor 
\begin{equation}\label{tensmet}
g_{\mu\nu}=\eta_{ab}{e^a}_\mu{e^{b}}_\nu ~ ,
\end{equation}
where $ \eta_{ab}=\text{diag}(-1,1,\dots,1) $ is the Minkowski metric in the Lorentz frame.

Additionally, the spacetime ${\mathcal M}^D$ can be classified by the Riemann curvature tensor which can be written in terms of the Christoffel symbol ${\Gamma^{\lambda}}_{\mu\nu} $,
\begin{eqnarray}\label{Riemanncurv}
{{\mathcal{R}}^{\rho}}_{\sigma\mu\nu}&=&\partial_{\mu}\Gamma^{\rho}_{~ \nu\sigma}-\partial_{\nu}\Gamma^{\rho}_{~ \mu\sigma}+\Gamma^{\rho}_{~ \mu\lambda}\Gamma^{\lambda}_{~ \nu\sigma}-\Gamma^{\rho}_{~ \nu\lambda}\Gamma^{\lambda}_{~ \mu\sigma} \notag\\
&=& {\bar{\mathcal{R}}^{\rho} }_{~ \sigma\mu\nu}+\nabla_\nu{\mathcal{K}^{\rho}}_{\mu\sigma}-\nabla_\mu{\mathcal{K}^{\rho}}_{\nu\sigma}+{\mathcal{K}^{\rho}}_{\nu\lambda}{\mathcal{K}^{\lambda}}_{\mu\sigma}-{\mathcal{K}^{\rho}}_{\mu\lambda}{\mathcal{K}^{\lambda}}_{\nu\sigma} ~ ,
\end{eqnarray} 
where
\begin{equation}\label{altRiemanncurv}
{\bar{\mathcal{R}}^{\rho}}_{~ \sigma\mu\nu}=\partial_{\mu}\omega^{\rho}_{~ \nu\sigma}-\partial_{\nu}\omega^{\rho}_{~ \mu\sigma}+\omega^{\rho}_{~ \mu\lambda}\omega^{\lambda}_{~ \nu\sigma}-\omega^{\rho}_{~ \nu\lambda}\omega^{\lambda}_{~ \mu\sigma} ~ ,
\end{equation}
with $ \nabla_\nu $ is the covariant derivative with respect to the spacetime metric \eqref{tensmet}. The Ricci tensor can be obtained by contracting \eqref{Riemanncurv} 
\begin{eqnarray}\label{Riccitens}
{\mathcal{R}}_{\mu\nu} = {\mathcal{R}}^{\rho}_{~ \mu\rho\nu} = {\bar{\mathcal{R}}}_{\mu\nu}+\nabla_\nu{\mathcal{K}^{\rho}}_{\rho\mu}-\nabla_\rho{\mathcal{K}^{\rho}}_{\nu\mu}+{\mathcal{K}^{\rho}}_{\nu\lambda}{\mathcal{K}^{\lambda}}_{\rho\mu}-{\mathcal{K}^{\rho}}_{\rho\lambda}{\mathcal{K}^{\lambda}}_{\nu\mu} ~ .
\end{eqnarray}
The Ricci scalar has the form
\begin{eqnarray}\label{Riccisclr}
{\mathcal{R}} = g^{\mu\nu}{{\mathcal{R}}}_{\mu\nu} = -T+2\nabla_\mu {T_\nu}^{\nu\mu} ~ .
\end{eqnarray}
%

\subsection{Equations of Motions}

Let us now shortly discuss the equations of motion in the scalar-torsion theory in which it contains the non-minimal derivative coupling term. The discussion in this subsection follows rather closely \cite{kofinas2012torsi}.

The action of the scalar-torsion theory with non-minimal derivative coupling has the form
 \begin{equation}\label{action}
S = -\frac{1}{2\kappa_D^2}\int d^D x ~ e T-\int d^D x e \Bigg[\left(\frac{1}{2}-\xi T\right)g^{\mu\nu}\partial_{\mu}\phi\partial_{\nu}\phi+V\Bigg] ~ ,
\end{equation}  
with $ \xi > 0$ is a coupling parameter whose root $ \sqrt{\xi } $ is a length scale in the theory and $e$ is defined as the vielbein determinant. The coupling $\kappa_D \equiv 1/M_p$ where $M_p$ is the D-dimensional Planck mass and $\xi > M_p$.

Varying \eqref{action} with respect to the vielbein, it gives
\begin{eqnarray}\label{vareaction}
\delta_e S &=&-\int d^D x\left(\frac{2}{\kappa_D^2}-4\xi\phi_{,\rho}\phi^{,\rho}\right)eS^{dca}\omega_{bdc}{e^{b}}_{\mu}\delta {e_a}^{\mu}-\int d^D x \Bigg[\Bigg(\frac{2}{\kappa_D}-4\xi\phi_{,\rho}\phi^{,\rho}\Bigg)e{S_a}^{\mu\nu}\delta {e^a}_{\nu}\Bigg]_{,\mu}\notag\\
&&+\int d^D x\Bigg\lbrace\left(\frac{2}{\kappa_D^2}-4\xi\phi_{,\rho}\phi^{,\rho}\right)\Bigg[(eS_\kappa^{\lambda}\nu{e_b}^{\kappa})_{,\nu}{e^{b}}_\mu+e\Bigg(\frac{1}{4}T\delta^{\lambda}_{\mu}-S^{\nu\kappa\lambda}T_{\nu\kappa\mu}\Bigg)\Bigg]\notag\\
&&+4\xi\Bigg[\frac{1}{2}eT\phi_{,\mu}\phi^{,\lambda}+e{S_{\mu}}^{\nu\lambda}(\phi_{,\kappa}\phi^{,\kappa})_{,\nu}\Bigg]+e\Bigg(\frac{1}{2}\phi_{,\rho}\phi^{,\rho}\delta^{\lambda}_{\mu}-\phi_{,\mu}\phi^{,\lambda}+V\delta^{\lambda}_{\mu}\Bigg)\Bigg\rbrace {e^a}_{\lambda}\delta {e_a}^\mu ~.\notag\\
\end{eqnarray} 
As argued in  \cite{kofinas2012torsi}, we could take a preferred frame $e_a$ such that the connection $ \omega_{bdc} $ vanishes and the boundary term is ignored. Setting $ \delta_eS=0 $, we obtain the torsion field equation
\begin{eqnarray}\label{torsioneom}
&&\left(\frac{2}{\kappa^2_D}-4\xi\phi_{,\rho}\phi^{,\rho}\right) \left[(e{S_\kappa}^{\lambda\nu} {e_{\bar{b}}}^{\kappa}),_{\nu}{e^{\bar{b}}}_{\mu}+e\left(\frac{1}{4}T\delta^\lambda_\mu-S^{\nu\kappa\lambda}T_{\nu\kappa\mu} \right)  \right]\notag\\
&&+4\xi\left[\frac{1}{2}eT\phi_{,\mu}\phi^{,\lambda}+e{S_\mu}^{\nu\lambda}(\phi_{,\kappa}\phi^{,\kappa}),\nu \right]
+e\left(\frac{1}{2}\phi_{,\rho}\phi^{,\rho}\delta^\lambda_{\mu}-\phi_{,\mu}\phi^{,\lambda}+V\delta^{\lambda}_{\mu}\right)=0 ~ ,\notag \\
\end{eqnarray}
which is an analog of the Einstein equation of motion in the standard gravitational theory. Next, we vary \eqref{action} with respect to the scalar field $ \phi $ such that we have
\begin{equation}\label{varscalaraction}
\delta_{\phi} S=\int d^Dx \left( \left[e(1-2\xi T)\phi^{,\mu} \right]_{,\mu}-e\frac{dV}{d\phi} \right) \delta\phi - \int d^Dx\left[e(1-2\xi T)\phi^{,\mu} \delta\phi \right]_{,\mu} ~ .
\end{equation}
Ignoring the boundary term and setting $ \delta_eS=0 $, we have
\begin{equation}\label{scalareom}
\left[e(1-2\xi T)\phi^{,\mu} \right]_{,\mu}-e\frac{dV}{d\phi}=0 ~ , 
\end{equation}
which is the scalar field equation of motion (the Klein-Gordon equation) in the theory.

As we have seen in \eqref{action}, this theory is indeed not a theory based on the curvatures, namely Riemann curvature, Ricci tensor, and Ricci scalar, which implies  that both \eqref{torsioneom} and \eqref{scalareom} do not contain such curvatures of the spacetime. Nevertheless, the solutions of \eqref{torsioneom} and \eqref{scalareom} describe a spacetime geometry with metric tensor \eqref{tensmet} which might have different pictures from the standard general relativity. As an alternative theory of gravity, it is of interest to study the simplest class of solutions, that is, the  static spacetimes. In the next section we will study these special class of solutions and check whether these solutions might be thought of as black holes. If these are physical, then it is sufficient to show the existence of horizon. In addition, the asymptotic solutions will be presented and compared to the result in \cite{Yaqin2017comment}.

\section{Static Spacetimes: General Setup}
\label{sec:staticgenset}

In this section, we will particularly discuss the static solutions of both \eqref{torsioneom} and \eqref{scalareom}. The starting point is to consider an ansatz metric of ${\mathcal M}^D$ given by
\begin{equation}\label{ansmet}
ds^2=-N(r)^2dt^2+K(r)^{-2}dr^2+R(r)^2d\Omega^2_{(D-2)} ~ ,
\end{equation}
where
\begin{equation}\label{ansmetsubman}
d\Omega^2_{(D-2)}=\hat{g}_{ij}(u)du^i du^j ~ ,
\end{equation} 
is a metric defined on $(D-2 )$-dimensional submanifold ${\mathcal S}^{D-2}$ spanned by the coordinates $u^i$ with $ \hat{g}^{ij}\hat{g}_{jk}=\delta^i_k $ and $i,j,k = 2, ..., D$. 
The metric functions $ N(r) $, $ K(r) $ and $ R(r) $ depend on the radial coordinate $ r $ such that the vielbein of the metric \eqref{ansmet} has the form
\begin{equation}\label{vielbeinD}
{e^a}_\mu=(N(r),K(r)^{-1},R(r){\hat{e}^{\bar{b}}}_i) ~ ,
\end{equation}
where $ {\hat{e}^{\bar{b}}}_i $ is the $(D-2)$-dimensional vielbein of ${\mathcal S}^{D-2}$. The determinant of the vielbein \eqref{vielbeinD} is
\begin{equation}\label{detvielbeinD}
e=\sqrt{ -g }=\frac{NR^{D-2}\sqrt{\hat{g}}}{K} ~ . 
\end{equation}
\subsection{Torsion and Curvature}
Inserting the ansatz \eqref{ansmet} to torsion scalar \eqref{torsionscalar}, we obtain
\begin{eqnarray}\label{torsionscalarD}
T &=& -(D-2)K^2\frac{R'}{R}\left[(D-3)\frac{R'}{R}+2\frac{N'}{N} \right]+\frac{\hat{T}}{R^2} ~ ,
\end{eqnarray}
where 
\begin{eqnarray}\label{torsionscalarD-2}
\hat{T}&=&\frac{1}{4}\hat{g}_{il}\hat{g}^{jm}\hat{g}^{kn}\hat{e}_{\bar{b}}^i\hat{e}_{\hat{c}}^l(\partial_n\hat{e}^{\bar{b}}_m-\partial_m\hat{e}^{\bar{b}}_n)(\partial_k\hat{e}^{\bar{c}}_j-\partial_j\hat{e}^{\bar{c}}_k)\notag\\
&&+\frac{1}{2}\hat{g}^{kl}\hat{e}_{\bar{b}}^i\hat{e}_{\hat{c}}^j(\partial_j\hat{e}^{\bar{b}}_l-\partial_l\hat{e}^{\bar{b}}_j)(\partial_i\hat{e}^{\bar{c}}_k-\partial_k\hat{e}^{\bar{c}}_i)\notag\\
&&-\hat{g}^{kl}\hat{e}_{\bar{b}}^i\hat{e}_{\hat{c}}^j(\partial_l\hat{e}^{\bar{b}}_i-\partial_i\hat{e}^{\bar{b}}_l)(\partial_k\hat{e}^{\bar{c}}_j-\partial_j\hat{e}^{\bar{c}}_k) ~ ,
\end{eqnarray}
is the torsion scalar on $(D-2)$-dimensional submanifold  ${\mathcal S}^{D-2}$ and for any function $f$, $f' \equiv df/dr$. Then, the quantity $\nabla_\mu {T_\nu}^{\nu\mu}$ in this case simplifies to
\begin{eqnarray}\label{divtorsion}
\nabla_\mu {T_\nu}^{\nu\mu} &=& -K^2\Bigg[\frac{K'N'}{KN}+\frac{N''}{N}+(D-2)\frac{K'R'}{KR}+(D-2)\frac{R''}{R}+(D-2)\frac{N'R'}{NR}\Bigg]-\notag\\
&&(D-2)K^2\frac{R'}{R}\left[(D-3)\frac{R'}{R}+2\frac{N'}{N} \right]+\notag\\
&&\frac{1}{2R^2}\Bigg[-\frac{2\mathcal{E}^i_{,i}}{\sqrt{\hat{g}}}+\hat{g}^{im}\hat{g}^{kn}{\hat{e}_{\bar{b}}}^l(\partial_{i}\hat{g}_{kl}+\partial_{k}\hat{g}_{li}-\partial_{l}\hat{g}_{ik})(\partial_{n}\hat{e}^{\bar{b}}_{m}-\partial_{m}\hat{e}^{\bar{b}}_{n})-\notag\\
&&\qquad\qquad\hat{g}^{im}\hat{g}^{jn}{\hat{e}_{\bar{b}}}^l(\partial_{i}\hat{g}_{jl}+\partial_{j}\hat{g}_{li}-\partial_{l}\hat{g}_{ij})(\partial_{n}\hat{e}^{\bar{b}}_{m}-\partial_{m}\hat{e}^{\bar{b}}_{n})+\notag\\
&&\qquad\qquad\hat{g}^{il}\hat{g}^{km}{\hat{e}_{\bar{b}}}^n(\partial_{i}\hat{g}_{kl}+\partial_{k}\hat{g}_{li}-\partial_{l}\hat{g}_{ik})(\partial_{n}\hat{e}^{\bar{b}}_{m}-\partial_{m}\hat{e}^{\bar{b}}_{n})\Bigg] ~ ,
\end{eqnarray}
where
\begin{eqnarray}\label{mathcalEtens}
\mathcal{E}^{i}&=&\sqrt{\hat{g}}\left(\hat{g}^{ij}\hat{e}_{\bar{b}}^k(\partial_j\hat{e}^{\bar{b}}_k-\partial_k\hat{e}^{\bar{b}}_j) \right).
\end{eqnarray}
The components of the Riemann curvature tensor in \eqref{Riemanncurv} in this case has the form
\begin{eqnarray}\label{Riemcomp}
	{\mathcal{R}^0}_{110}=-{\mathcal{R}^0}_{101}&=&\frac{N''}{N}+\frac{N'K'}{NK} ~ , \notag\\
{\mathcal{R}^1}_{010}=-{\mathcal{R}^1}_{001}&=&NN''K^2+NN'KK'  ~ , \notag\\
{\mathcal{R}^0}_{ij0}=-{\mathcal{R}^0}_{i0j}&=&\frac{N'R'K^2R}{N}\hat{g}_{ij}  ~ , \notag\\
{\mathcal{R}^i}_{0j0}=-{\mathcal{R}^i}_{00j}&=&\frac{N'R'K^2N}{R}{\delta^i}_{j}  ~ , \notag\\
{\mathcal{R}^1}_{i1j}=-{\mathcal{R}^1}_{ij1}&=&-\left( RR''K^2+RR'KK'-\left( D-3 \right){{K}^{2}}{{{R}'}^{2}}\right)\hat{g}_{ij}  ~ ,  \notag\\
{\mathcal{R}^i}_{11j}=-{\mathcal{R}^i}_{1j1}&=&\left( \frac{R''}{R}+\frac{R'K'}{RK}\right){\delta^i}_j  ~ ,  \notag\\
{\mathcal{R}}^i_{jkl}&=&{\hat{\mathcal{R}}^i}_{jkl} ~ ,
\end{eqnarray} 
where
\begin{equation}\label{RiemcompD-2}
{\hat{\mathcal{R}}^i}_{jkl} = \hat{\bar{\mathcal{R}}}^{i}_{jkl}+\nabla_l{\mathcal{K}^{i}}_{kj}-\nabla_k{\mathcal{K}^{i}}_{lj}+{\mathcal{K}^{i}}_{lm}{\mathcal{K}^{m}}_{kj}-{\mathcal{K}^{i}}_{km}{\mathcal{K}^{m}}_{lj} ~ .
\end{equation}
Then, the norm of the Riemann curvature tensor called Kretschmann scalar is given by
\begin{eqnarray}\label{NormRiem}
\|Riem\|^2 &\equiv& \mathcal{R}^{\rho\sigma\mu\nu}\mathcal{R}_{\rho\sigma\mu\nu}\notag\\
	&=&\frac{{{\widehat{\mathcal{R}}}^{ijkl}}{{\widehat{\mathcal{R}}}_{ijkl}}}{{{R}^{4}}}+4{{\left( \frac{{N}''{{K}^{2}}}{N}+\frac{{N}'K{K}'}{N} \right)}^{2}}+4(D-2)\frac{{{{{N}'}}^{2}}{{{{R}'}}^{2}}{{K}^{4}}}{{{N}^{2}}{{R}^{2}}} \notag\\ 
	&&+4(D-2){{\left( \frac{{R}''{{K}^{2}}}{R}+\frac{{R}'K{K}'}{R}-\left( D-3 \right)\frac{{{K}^{2}}{{{{R}'}}^{2}}}{{{R}^{2}}} \right)}^{2}} ~ ,
\end{eqnarray}
which will be useful for our analysis in the next section.

The components of the spacetime Ricci tensor in this case are
\begin{eqnarray}\label{compRiccitens}
{\mathcal{R}}_{00}&=&N^2K^2\left( \frac{N''}{N}+\frac{N' K'}{NK} + (D-2)\frac{N'R'}{NR}\right) ~ , \notag\\
{\mathcal{R}}_{11}&=&-\left( \frac{N''}{N}+\frac{N'K'}{NK}+(D-2)\frac{R''}{R}+(D-2)\frac{R'K'}{RK}\right) ~ ,\notag\\
{{\mathcal{R}}_{ij}}&=&{{\hat{\mathcal{R}}}_{ij}}-{{R}^{2}}{{K}^{2}}\left( \frac{{{R}''}}{R}+\frac{{R}'{K}'}{RK}+\frac{{N}'{R}'}{NR} \right){{\hat{g}}_{ij}}-{{K}^{2}}{{{R}'}^{2}}{{\hat{g}}_{ij}}\left( D-3 \right) ~ ,
\end{eqnarray}
where
\begin{equation}\label{compRiccitensSD}
\hat{\mathcal{R}}_{ij}=\hat{\bar{\mathcal{R}}}_{ij}+\nabla_j{\mathcal{K}^{k}}_{k i}-\nabla_k{\mathcal{K}^{k}}_{ji}+{\mathcal{K}^{k}}_{jl}{\mathcal{K}^{l}}_{k i}-{\mathcal{K}^{k}}_{kl}{\mathcal{K}^{l}}_{ji} ~ ,
\end{equation}
are the components of the Ricci tensor of the submanifold ${\mathcal S}^{D-2}$. Contracting the Ricci tensor given in \eqref{compRiccitens} by the metric \eqref{ansmet}, we get the Ricci scalar  
\begin{equation}\label{Riccisclr1}
\mathcal{R}=\frac{\widehat{\mathcal{R}}}{{{R}^{2}}}-2{{K}^{2}}\left[ \frac{{K}'{N}'}{KN}+\frac{{{N}''}}{N}+(D-2)\frac{{K}'{R}'}{KR}+(D-2)\frac{{{R}''}}{R}+(D-2)\frac{{N}'{R}'}{NR}+\frac{{{{{R}'}}^{2}}}{{{R}^{2}}}\frac{\left( D-3 \right)\left( D-2 \right)}{2} \right] ~ ,
\end{equation}
where
\begin{eqnarray}\label{RiccisclrD-2}
\hat{\mathcal{R}} &=& -2\left(\frac{\hat{T}}{2}+ \frac{\mathcal{E}^i_{,i}}{\sqrt{\hat{g}}}\right) +\hat{g}^{im}\hat{g}^{kn}{\hat{e}_{\bar{b}}}^l(\partial_{i}\hat{g}_{kl}+\partial_{k}\hat{g}_{li}-\partial_{l}\hat{g}_{ik})(\partial_{n}\hat{e}^{\bar{b}}_{m}-\partial_{m}\hat{e}^{\bar{b}}_{n}) \notag\\
&& -  \hat{g}^{im}\hat{g}^{jn}{\hat{e}_{\bar{b}}}^l(\partial_{i}\hat{g}_{jl}+\partial_{j}\hat{g}_{li}-\partial_{l}\hat{g}_{ij})(\partial_{n}\hat{e}^{\bar{b}}_{m}-\partial_{m}\hat{e}^{\bar{b}}_{n}) \notag\\
&& +  \hat{g}^{il}\hat{g}^{km}{\hat{e}_{\bar{b}}}^n(\partial_{i}\hat{g}_{kl}+\partial_{k}\hat{g}_{li}-\partial_{l}\hat{g}_{ik})(\partial_{n}\hat{e}^{\bar{b}}_{m}-\partial_{m}\hat{e}^{\bar{b}}_{n})  ~ ,
\end{eqnarray}
is the Ricci scalar of ${\mathcal S}^{D-2}$. In this paper we assume that all ${\hat{\mathcal{R}}^i}_{jkl}$, $\hat{\mathcal{R}}_{ij}$ and $\hat{\mathcal{R}}$ are finite for all $i, j, k, l$.



%
\subsection{Master equation}
Now, we are ready to discuss the static solutions of both the torsion field equation \eqref{torsioneom} and the Klein Gordon equation \eqref{scalareom}. Inserting the vielbein \eqref{vielbeinD} and the torsion scalar \eqref{torsionscalarD} into \eqref{torsioneom} and assuming $\phi(r)$, we find some equations as the following

\begin{enumerate}
	\item For $ \mu=0 $ and $ \lambda=0 $
		\begin{eqnarray}\label{toreq1}
		&&2\left(\frac{1}{\kappa^2_D K^2}-2\xi\phi'^2\right)\Bigg[(D-2)\frac{R'K'}{RK}+\frac{(D-2)(D-3)}{2}\frac{R'^2}{R^2}+(D-2)\frac{R''}{R}+\frac{\hat{T}}{2R^2K^2} \notag\\
		&& +\frac{{\mathcal{E}^{i}}_{,i}}{\sqrt{\hat{g}}R^2K^2}\Bigg] - 8(D-2)\xi\phi'\frac{R'}{R}\left(\frac{K'}{K}\phi'+\phi'' \right)+\frac{1}{K^2}\left( \phi'^2+\frac{2V}{K^2}\right) = 0 ~ .
		\end{eqnarray}
	\item For $ \mu=1 $ and $ \lambda=1 $
		\begin{eqnarray}\label{toreq2}
		&&2\left(2\xi\phi'^2-\frac{1}{\kappa^2_D K^2}\right)\left[\frac{R'}{R}\left(\frac{(D-2)(D-3)}{2}\frac{R'}{R}+(D-2)\frac{N'}{N} \right) +\frac{\hat{T}}{2R^2K^2}+\frac{{\mathcal{E}^{i}}_{,i}}{\sqrt{\hat{g}}R^2K^2}  \right] \notag\\
		&&+4(D-2)\xi\phi'^2\left( \frac{R'}{R}\left((D-3)\frac{R'}{R}+\frac{2N'}{N}\right) -\frac{\hat{T}}{(D-2)R^2K^2}\right) +\frac{1}{K^2}\left( \phi'^2-\frac{2V}{K^2}\right) =0 ~ .\notag\\
		\end{eqnarray}
	\item For $ \mu=i $ and $ \lambda=i $
		\begin{eqnarray}\label{toreq3}
		&&2\left(\frac{1}{\kappa^2_D K^2}-2\xi\phi'^2\right)\Bigg[\frac{N'}{N}\left((D-3)\frac{R'}{R}+\frac{K'}{K} \right)+(D-3)\left( \frac{R''}{R}+\frac{R'K'}{RK}\right)+\frac{N''}{N}  \notag\\
		&& + \frac{(D-3)(D-4)}{2}\frac{R'^2}{R^2}+\frac{2}{(D-2)R^2K^2}\left( \frac{(D-6)}{4}\hat{T}+\frac{{{\hat{e}}^{\bar{b}}}_{i}\left( {\mathcal{F}^{ik}_{\bar{b}}}\right) _{,k}}{\sqrt{\hat{g}}}\right)\Bigg] \notag\\
		&&-8\xi\phi'\left((D-3)\frac{R'}{R}+\frac{N'}{N}\right) \left(\frac{K'}{K}\phi'+\phi'' \right)+\frac{1}{K^2}\left( \phi'^2+\frac{2V}{K^2}\right) = 0 ~ ,
		\end{eqnarray}
	where  
	\begin{eqnarray}\label{mathcalF}
	{\mathcal{F}^{ik}_{\bar{b}}}&=&\sqrt{\hat{g}}{\hat{S}_j}^{ik}{\hat{e}^j}_{\bar{b}}\notag\\
	&=&\sqrt{\hat{g}}\Bigg[\frac{1}{2}\Big[-\hat{g}^{il}{{\hat{e}}_{\bar{b}}}^{k}{{\hat{e}}_{\bar{b}}}^{m}\left(\partial_m\hat{e}^{\bar{b}}_l-\partial_l\hat{e}^{\bar{b}}_m\right)+\hat{g}^{il}{{\hat{e}}_{\bar{b}}}^{j}\hat{e}_{\bar{c}}^k\left(\partial_l\hat{e}^{\bar{c}}_j-\partial_j\hat{e}^{\bar{c}}_l\right)\notag\\
	&&+\hat{g}^{km}{{\hat{e}}_{\bar{b}}}^{j}\hat{e}_{\bar{c}}^i\left(\partial_j\hat{e}^{\bar{c}}_m-\partial_m\hat{e}^{\bar{c}}_j\right)\Big]-\hat{g}^{im}{{\hat{e}}_{\bar{b}}}^{k}\hat{e}_{\bar{c}}^l\left(\partial_m\hat{e}^{\bar{c}}_l-\partial_l\hat{e}^{\bar{c}}_m\right)+\hat{g}^{km}{{\hat{e}}_{\bar{b}}}^{i}\hat{e}_{\bar{c}}^l\left(\partial_m\hat{e}^{\bar{c}}_l-\partial_l\hat{e}^{\bar{c}}_m\right)\Bigg] ~ . \notag\\
	\end{eqnarray}
	\item For $ \mu=1 $ and $ \lambda=i $
	\begin{equation}\label{toreq4}
	\left(\phi'^2-\frac{1}{2\kappa^2_D K^2\xi}\right)\frac{(D-4)R'}{R}+2\phi'\left(\frac{K'}{K}\phi'+\phi''\right) = 0 ~ .
	\end{equation}
\end{enumerate}
 The Klein-Gordon equations \eqref{scalareom} in this case become
\begin{eqnarray}\label{KGstatic}
&&\left\lbrace NR^{D-2}K\left[1+2(D-2)\xi\left(\frac{K^2R'}{R}\left[(D-3)\frac{R'}{R}+\frac{2N'}{N} \right]+\frac{\hat{T}}{R^2} \right) \right]\phi' \right\rbrace'\notag\\
&&\qquad\qquad -\frac{NR^{D-2}}{K}\frac{dV}{d\phi} = 0 ~ .
\end{eqnarray} 
It is worth to mention that $ \hat{T} $, $ \mathcal{E}^{i} $, $ {\mathcal{F}}^{ik}_{\bar{b}} $, and $\hat{g}$ depend on the local coordinate $u^i$ of the submanifold ${\mathcal S}^{D-2}$. As we immediately see from \eqref{toreq1}-\eqref{toreq3} and \eqref{KGstatic}, the triple quantities $\hat{T}$, $\frac{ {\mathcal{E}^{i}}_{,i}}{\sqrt{\hat{g}}}$, and  $\frac{{{\hat{e}}^{\bar{b}}}_{i}\left( {\mathcal{F}^{ik}_{\bar{b}}}\right) _{,k}}{\sqrt{\hat{g}}}$ should be constants (see below). In the theory, we have four dynamical functions, namely $K(r)$, $N(r)$, $\phi(r)$, and $V(r)$ since the form of the function $R(r)$ can be chosen by the radial coordinate redefinition. Therefore, among  \eqref{toreq1}-\eqref{toreq3}, \eqref{toreq4}, and  \eqref{KGstatic}, one of them is expected to be the constraint of the theory. That is \eqref{toreq4}, which can be simplified into
\begin{equation}\label{constraint}
\phi'=\frac{1}{K}\left(\frac{1}{2\kappa_D^2\xi}+\frac{C}{R^{D-4}}\right)^{\frac{1}{2}}  ~ ,
\end{equation}
where $ \phi'  = d\phi / dr$ and $ C $ is the integration constant chosen to be $C \ge 0$. It is worth to mention  that  for the case of  $C=0$ we have a special class of exact solutions for $D >  4$ with constant scalar potential $V(\phi)$ as discussed in section \ref{sec:specsolexact}. However, we only use the assumption $C > 0$ in the construction of global existence discussed in section \ref{sec:localglobalex} since it diverges for $C=0$. The case of $C < 0$ and $D >4$ is also excluded because it leads to a contradiction with the standard transformation of the static metric \eqref{ansmet}. This will be discussed in section \ref{sec:propertiesstatic}.

Then, we insert \eqref{constraint} into  \eqref{toreq1}-\eqref{toreq3} and \eqref{KGstatic}, and introducing $\dot{f} \equiv df/d\phi$, equations \eqref{toreq1}-\eqref{toreq3} and \eqref{KGstatic} can be cast into the following form:
%
\begin{eqnarray}\label{toreq5}
	&& \frac{{\ddot{R}}}{R}-\left[ \frac{(D-5)}{2}+\frac{\kappa _{D}^{2}\xi C\left( D-4 \right)}{\left( {{R}^{D-4}}+2\kappa _{D}^{2}\xi C \right)} \right]\frac{{{{\dot{R}}}^{2}}}{{{R}^{2}}}+\frac{2\kappa _{D}^{2}\xi {{R}^{(D-4)}}}{(D-2)\left( {{R}^{(D-4)}}+2\kappa _{D}^{2}\xi C \right)}\notag\\
	&& \times \left[ \frac{1}{{{R}^{2}}}\left( \frac{{\hat{T}}}{2}+\frac{1}{\sqrt{{\hat{g}}}}{{\mathcal{E}}^{i}}_{,i} \right)-\frac{{{R}^{(D-4)}}}{4\xi C}\left( \frac{1}{2\kappa _{D}^{2}\xi }+\frac{C}{{{R}^{D-4}}}+2V \right) \right]=0 ~ , \notag\\
\end{eqnarray}
\begin{eqnarray}\label{toreq6}
	&& \frac{{\dot{R}}}{R}\left[ \frac{(D-2)(D-3)}{2}\left( 1+\frac{{{R}^{D-4}}}{\kappa _{D}^{2}\xi C} \right)\frac{{\dot{R}}}{R}+(D-2)\left( 1+\frac{{{R}^{D-4}}}{\kappa _{D}^{2}\xi C} \right)\frac{{\dot{N}}}{N} \right]-\frac{2{{R}^{D-4}}}{C\left( {{R}^{D-4}}+2\kappa _{D}^{2}\xi C \right)} \notag\\
	&& \times \left[ \frac{{\hat{T}}}{2{{R}^{2}}}\left( {{R}^{D-4}}+3\kappa _{D}^{2}\xi C \right)+\left( \kappa _{D}^{2}\xi C \right)\frac{{{\mathcal{E}}^{i}}_{,i}}{\sqrt{{\hat{g}}}{{R}^{2}}}-\frac{\kappa _{D}^{2}{{R}^{D-4}}}{4}\left( \frac{1}{2\kappa _{D}^{2}\xi }+\frac{C}{{{R}^{D-4}}}-2V \right) \right]=0 ~ , \notag\\
\end{eqnarray}
\begin{eqnarray}\label{toreq7}
	&& \frac{{\ddot{N}}}{N}-\left[ \frac{(D-3)}{2} \right]\frac{{{{\dot{R}}}^{2}}}{{{R}^{2}}}+\left[ 1-\frac{\left( D-4 \right)\kappa _{D}^{2}\xi C}{\left( {{R}^{D-4}}+2\kappa _{D}^{2}\xi C \right)} \right]\frac{{\dot{N}}}{N}\frac{{\dot{R}}}{R}+\frac{2\kappa _{D}^{2}\xi {{R}^{D-4}}}{(D-2)\left( {{R}^{D-4}}+2\kappa _{D}^{2}\xi C \right){{R}^{2}}}\notag\\
	&& \times \left( -\frac{3}{2}\hat{T}+2\frac{{{{\hat{e}}}^{{\bar{b}}}}_{i}{{\left( \mathcal{F}_{{\bar{b}}}^{ik} \right)}_{,k}}}{\sqrt{{\hat{g}}}}-(D-3)\frac{{{\mathcal{E}}^{i}}_{,i}}{\sqrt{{\hat{g}}}}-\frac{{{R}^{D-2}}}{4C\xi }\left( \frac{1}{2\kappa _{D}^{2}\xi }+\frac{C}{{{R}^{D-4}}}+2V \right) \right)=0 ~ ,\notag\\
\end{eqnarray}
\begin{eqnarray}\label{KGstatik1}
&&\Bigg\lbrace NR^{D-2}\left(\frac{1}{2\kappa_D^2\xi}+\frac{C}{R^{D-4}}\right)^{\frac{1}{2}}\Bigg[1+2(D-2)\xi\left(\frac{1}{2\kappa_D^2\xi}+\frac{C}{R^{D-4}}\right) \notag\\
&& \times \left(\frac{\dot{R}}{R}\left[(D-3)\frac{\dot{R}}{R}+\frac{2\dot{N}}{N} \right]+\frac{\hat{T}}{R^2} \right) \Bigg]\Bigg\rbrace^{\cdot}-\frac{NR^{D-2}}{\left(\frac{1}{2\kappa_D^2\xi}+\frac{C}{R^{D-4}}\right)^{\frac{1}{2}}}\dot{V} = 0 ~ .
\end{eqnarray}
%
It is important to notice that \eqref{toreq6} is again a constraint since it does not contain the second order derivative. After some computation using \eqref{toreq5}- \eqref{toreq7}, we can rederive \eqref{KGstatik1}, which shows that \eqref{KGstatik1} is redundant. Thus, it is sufficient to consider only three equations.

Next, we define new variables as in \cite{kofinas2012torsi} with slightly modification
\begin{equation}\label{Y40}
Y\equiv y^2 ~ ,
\end{equation}
\begin{equation}\label{Z41}
 Z\equiv \frac{z^2}{\zeta} ~ ,
\end{equation}
where $\zeta$ is a non-zero constant and $ x,y,z $ 
\begin{equation}\label{x42}
x=\ln R ~ ,
\end{equation}
\begin{equation}\label{y43}
y=\frac{\dot{R}}{R} ~ ,
\end{equation}
\begin{equation}\label{z44}
z=\frac{{(RN^2)}^{\cdot}}{RN^2} ~ ,
\end{equation}
so that we have
\begin{equation}\label{45}
\dot{y}=\frac{1}{2}\frac{dY}{dx} ~ ,
\end{equation}
\begin{equation}\label{46}
\dot{z}=\frac{1}{2}\zeta\sqrt{\frac{Y}{Z}}\frac{dZ}{dx} ~ .
\end{equation}
%
Then, \eqref{toreq5}-\eqref{toreq7} become simply
\begin{eqnarray}\label{toreq8}
	&& \frac{dY}{dx}-\left[ (D-7)+\frac{2\kappa _{D}^{2}\xi C\left( D-4 \right)}{\left( {{e}^{(D-4)x}}+2\kappa _{D}^{2}\xi C \right)} \right]Y+\frac{4\kappa _{D}^{2}\xi {{e}^{(D-6)x}}}{(D-2)({{e}^{(D-4)x}}+2\kappa _{D}^{2}\xi C)}\notag\\ 
	&& \times \left[ \left( \frac{{{\lambda }_{1}}}{2}+{{\lambda }_{2}} \right)-\frac{{{e}^{(D-2)x}}}{4\xi C}\left( \frac{1}{2\kappa _{D}^{2}\xi }+\frac{C}{{{e}^{(D-4)x}}}+2V \right) \right]=0 \notag\\  ~ ,
\end{eqnarray}
\begin{eqnarray}\label{toreq9}
	&&  \sqrt{\zeta YZ}+\left(D-4\right)Y-\frac{4\kappa _{D}^{2}\xi {{e}^{\left( D-6 \right)x}}}{(D-2)\left( {{e}^{\left( D-4 \right)x}}+\kappa _{D}^{2}\xi C \right)\left( {{e}^{\left( D-4 \right)x}}+2\kappa _{D}^{2}\xi C \right)}\notag\\ 
	&& \times \left[ \frac{{{\lambda }_{1}}}{2}\left( {{e}^{\left( D-4 \right)x}}+3\kappa _{D}^{2}\xi C \right)+\left( \kappa _{D}^{2}\xi C \right){{\lambda }_{2}}-\frac{\kappa _{D}^{2}{{e}^{\left( D-2 \right)x}}}{4}\left( \frac{1}{2\kappa _{D}^{2}\xi }+\frac{C}{{{e}^{\left( D-4 \right)x}}}-2V \right) \right]=0 ~ ,\notag\\
\end{eqnarray}
\begin{eqnarray}\label{toreq10}
		&& \zeta \sqrt{\frac{Y}{Z}}\frac{dZ}{dx}-3\left( D-4 \right)Y+{{\zeta }}Z-\sqrt{\zeta YZ}\frac{2\left( D-4 \right)\kappa _{D}^{2}\xi C}{\left( {{e}^{(D-4)x}}+2\kappa _{D}^{2}\xi C \right)}+\frac{4\kappa _{D}^{2}\xi {{e}^{(D-6)x}}}{({{e}^{(D-4)x}}+2\kappa _{D}^{2}\xi C)\left( D-2 \right)} \notag\\ 
		&& \times \left[ -\frac{5{{\lambda }_{1}}}{2}+4{{\lambda }_{3}}-2(D-7){{\lambda }_{2}}-\frac{3{{e}^{(D-2)x}}}{4C\xi }\left( \frac{1}{2\kappa _{D}^{2}\xi }+\frac{C}{{{e}^{(D-4)x}}}+2V \right) \right]=0 ~ ,
\end{eqnarray}
where we have defined
\begin{eqnarray}\label{torsionconstants}
\hat{T} &\equiv& \lambda_1 ~ , \notag\\
\frac{ {\mathcal{E}^{i}}_{,i}}{\sqrt{\hat{g}}}  &\equiv&  \lambda_2  ~ , \notag\\
\frac{{{\hat{e}}^{\bar{b}}}_{i}\left( {\mathcal{F}^{ik}_{\bar{b}}}\right) _{,k}}{\sqrt{\hat{g}}}  &\equiv& \lambda_3 ~ .
\end{eqnarray}
%

\begin{lemma}
	\label{3lambdakons}
$ \lambda_1 $, $ \lambda_2 $ and $ \lambda_2 $ are real constants.
\end{lemma}
\begin{proof}
Let us first consider \eqref{toreq8} which can be rewritten as
\begin{eqnarray}\label{toreq81}
 \frac{\lambda_1}{2}+\lambda_2   &=& \frac{e^{(D-2)x}}{4\xi C}\left(\frac{1}{2\kappa_D^2\xi}+\frac{C}{e^{(D-4)x}}+2V\right)\notag\\
&&  +  \frac{ (D-2)(e^{(D-4)x}+2\kappa_D^2\xi C)}{ 4\kappa_D^2\xi e^{(D-6)x}}  \left[ \frac{dY}{dx}-\left( {(D-7)}+\frac{2(D-4)\kappa_D^2\xi}{\left(e^{(D-4)x}+2\kappa_D^2\xi C\right)}\right)Y \right] ~ .\notag\\
\end{eqnarray}
It is easy to see that since the left hand side of \eqref{toreq81} depends only on $u$, while the right hand side of \eqref{toreq81} depends only on the radial coordinate $r$, we obtain that $ \frac{\lambda_1}{2}+\lambda_2$ has to be a real constant. Then, we employ similar analysis for \eqref{toreq9} and \eqref{toreq10}, we conclude that $ \lambda_1 $, $ \lambda_2 $ and $ \lambda_2 $ are real constants.
\end{proof}


After some computation, we could show that \eqref{toreq8}-\eqref{toreq10} can be simplified into a highly nonlinear ordinary differential equation which we will call the master equation,
{\footnotesize{
	 \begin{eqnarray}\label{mastereq}
		&& \frac{{{d}^{2}}Y}{d{{x}^{2}}}-\frac{1}{2}{{\left( \frac{dY}{dx} \right)}^{2}}+\frac{dY}{dx}\left\{ \left( \frac{\left( D-4 \right)}{{\tilde{\eta }}}-\frac{4\kappa _{D}^{2}\xi C}{\eta } \right){{e}^{\left( D-4 \right)x}}+\left( 8\kappa _{D}^{2}{{\xi }^{2}}{{C}^{2}}-\frac{4(D-2)}{\kappa _{D}^{2}} \right)(D-4)\tilde{\eta } \right.+4\xi C(D-7) \notag\\ 
		&& \left. -\frac{4\xi {{\lambda }_{1}}\hat{\eta }+8\kappa _{D}^{2}{{\xi }^{2}}C{{\lambda }_{2}}}{\eta }{{e}^{\left( D-6 \right)x}}+\left( \frac{2}{\eta }+\frac{2\kappa _{D}^{2}\xi }{\eta }\left( \frac{{{\lambda }_{1}}}{2}+{{\lambda }_{2}} \right) \right){{e}^{\left( 2D-8 \right)x}} \right\}-\frac{6(D-2)\left( D-4 \right)}{{{\zeta }^{1/2}}\kappa _{D}^{2}}\tilde{\eta }{{Y}^{2}} \notag\\ 
		&& -\left\{ 2\xi C(D-7)+\frac{4\kappa _{D}^{2}{{\xi }^{2}}{{C}^{2}}\left( D-4 \right)}{\eta }+\frac{2(D-2)D-4)\tilde{\eta }}{\kappa _{D}^{2}} \right\}Y\frac{dY}{dx}-\frac{8\kappa _{D}^{2}\xi C\left( \tilde{\eta }+\eta  \right)\left( D-4 \right)}{\tilde{\eta }{{\eta }^{2}}}{{e}^{\left( 2D-8 \right)x}} \notag\\ 
		&& +\frac{8{{\lambda }_{1}}\xi \left( \tilde{\eta }\eta -\eta \hat{\eta }-\tilde{\eta }\hat{\eta } \right)-8\kappa _{D}^{2}{{\xi }^{2}}C{{\lambda }_{2}}\left( \eta +\tilde{\eta } \right)}{\tilde{\eta }{{\eta }^{2}}}\left( D-4 \right){{e}^{2\left( D-5 \right)x}}+\frac{8{{\lambda }_{1}}\xi \hat{\eta }+16\kappa _{D}^{2}{{\xi }^{2}}C{{\lambda }_{2}}}{\eta }\left( D-6 \right){{e}^{\left( D-6 \right)x}} \notag\\ 
		&& +\frac{8\kappa _{D}^{2}\xi C\left( D-4 \right)}{\eta }{{e}^{\left( D-4 \right)x}}-\left( \kappa _{D}^{2}+\kappa _{D}^{4}\xi \left( \frac{{{\lambda }_{1}}}{2}+{{\lambda }_{2}} \right) \right)\left( \frac{8\left( D-4 \right){{e}^{\left( 2D-8 \right)x}}}{\kappa _{D}^{2}\eta }-\frac{4\left( \tilde{\eta }+\eta  \right)\left( D-4 \right){{e}^{3\left( D-4 \right)x}}}{\kappa _{D}^{2}\tilde{\eta }{{\eta }^{2}}} \right) \notag\\ 
		&& -\left( \frac{8\kappa _{D}^{2}{{\xi }^{2}}{{C}^{2}}\left( D-4 \right)\left( \tilde{\eta }+\eta  \right)}{\eta }+\frac{(D-7)\xi C}{{\tilde{\eta }}} \right)\left( D-4 \right){{e}^{\left( D-4 \right)x}}Y+\frac{2(D-2)}{{{\zeta }^{1/2}}\kappa _{D}^{2}}\tilde{\eta }\left\{ \frac{2\kappa _{D}^{2}\xi {{\lambda }_{1}}\hat{\eta }+4\kappa _{D}^{4}{{\xi }^{2}}C{{\lambda }_{2}}}{(D-2)}\frac{{{e}^{\left( D-6 \right)x}}}{\tilde{\eta }\eta } \right. \notag\\ 
		&& -\left( \left( \frac{{{\lambda }_{1}}}{2}+{{\lambda }_{2}} \right)\frac{\kappa _{D}^{4}\xi }{(D-2)}+\frac{\kappa _{D}^{2}}{(D-2)} \right)\frac{{{e}^{\left( 2D-8 \right)x}}}{\tilde{\eta }\eta }+\frac{2C\kappa _{D}^{4}\xi }{(D-2)}\frac{{{e}^{\left( D-4 \right)x}}}{\tilde{\eta }\eta }+\frac{\kappa _{D}^{2}}{4(D-2)\tilde{\eta }}\frac{dY}{dx}+(D-7)\frac{\kappa _{D}^{2}\xi C}{(D-2)\tilde{\eta }}Y \notag\\ 
		&& +{{\left. \frac{2{{\left( \kappa _{D}^{2}\xi C \right)}^{2}}\left( D-4 \right)}{(D-2)\tilde{\eta }\eta }Y-(D-4)Y \right\}}^{2}}-\frac{4\left( D-4 \right)(D-2)\xi C}{{{\zeta }^{1/2}}\eta }Y\left\{ \frac{2\kappa _{D}^{2}\xi {{\lambda }_{1}}\hat{\eta }+4\kappa _{D}^{4}{{\xi }^{2}}C{{\lambda }_{2}}}{(D-2)}\frac{{{e}^{\left( D-6 \right)x}}}{\tilde{\eta }\eta } \right. \notag\\ 
		&& -\left( \left( \frac{{{\lambda }_{1}}}{2}+{{\lambda }_{2}} \right)\frac{\kappa _{D}^{4}\xi }{(D-2)}+\frac{\kappa _{D}^{2}}{(D-2)} \right)\frac{{{e}^{\left( 2D-8 \right)x}}}{\tilde{\eta }\eta }+\frac{2C\kappa _{D}^{4}\xi }{(D-2)}\frac{{{e}^{\left( D-4 \right)x}}}{\tilde{\eta }\eta }+\frac{\kappa _{D}^{2}}{4(D-2)\tilde{\eta }}\frac{dY}{dx}+(D-7)\frac{\kappa _{D}^{2}\xi C}{(D-2)\tilde{\eta }}Y \notag\\ 
		&& \left. +\frac{2{{\left( \kappa _{D}^{2}\xi C \right)}^{2}}\left( D-4 \right)}{(D-2)\tilde{\eta }\eta }Y-(D-4)Y \right\}+\frac{8\xi {{e}^{(D-6)x}}\tilde{\eta }}{\eta {{\zeta }^{1/2}}}Y\left\{ -\frac{5{{\lambda }_{1}}}{2}+4{{\lambda }_{3}}-2(D-7){{\lambda }_{2}}-\frac{\eta }{16\kappa _{D}^{2}C{{\xi }^{2}}{{e}^{(D-6)x}}}\frac{dY}{dx} \right. \notag\\ 
		&& +\left[ (D-7)+\frac{2\kappa _{D}^{2}\xi C\left( D-4 \right)}{\eta } \right]\frac{3\eta }{4\kappa _{D}^{2}\xi {{e}^{(D-6)x}}}Y\left. -\frac{3{{e}^{(D-2)x}}}{4C\xi }\left( \frac{{{\lambda }_{1}}}{2}+{{\lambda }_{2}} \right) \right\}=0 ~,	
	\end{eqnarray} } }
where 
\begin{eqnarray}
\eta(x) &=& e^{(D-4)x}+2\kappa_D^2\xi C ~ , \notag\\
\hat{\eta}(x) &=& e^{(D-4)x}+3\kappa_D^2\xi C ~ , \notag\\
\tilde{\eta}(x ) &=& e^{(D-4)x}+\kappa_D^2\xi C ~ .
\end{eqnarray}
This master equation \eqref{mastereq} describes the $D$-dimensional static spacetimes of scalar-torsion theories with non-minimal derivative coupling which generalizes the four dimensional case \cite{Yaqin2017comment}. 
If we could have the exact form of $Y$, then using \eqref{toreq8} and \eqref{toreq9} the function $Z$ could be extracted via
  \begin{eqnarray} \label{ZeqY}
 	Z&=&\frac{1}{\zeta Y}\left\{ -(D-4)Y+\frac{4\kappa _{D}^{2}\xi {{e}^{\left( D-6 \right)x}}}{(D-2)\tilde{\eta }\eta }\left\{ \frac{{{\lambda }_{1}}}{2}\hat{\eta }+\left( \kappa _{D}^{2}\xi C \right){{\lambda }_{2}} \right. \right. \notag\\ 
 	&+&\frac{C(D-2)\eta }{4{{e}^{(D-6)x}}}\frac{dY}{dx}+\left( \frac{{{\lambda }_{1}}}{2}+{{\lambda }_{2}} \right)\frac{\kappa _{D}^{2}{{e}^{\left( D-2 \right)x}}}{4}-\frac{\eta {{e}^{2x}}}{4\xi } \notag\\ 
 	&-& {{\left. \left. \left[ (D-7)\frac{C(D-2)\eta }{4{{e}^{(D-6)x}}}+\frac{{{C}^{2}}\kappa _{D}^{2}\xi (D-2)\left( D-4 \right)}{2{{e}^{(D-6)x}}} \right]Y \right\} \right\}}^{2}}~ .
 \end{eqnarray}

Let us now consider the spacetime metric \eqref{ansmet}. From equations (\ref{Y40}), (\ref{x42}) and (\ref{z44}), we have
 \begin{equation}\label{Y55}
 Y=\left(\frac{dx}{d\phi}\right)^2 ~ ,
 \end{equation}
and 
\begin{equation}\label{ZY56}
 \frac{Z}{Y}=\left(\frac{d\ln(RN^2)}{dx} \right)^2 ~ ,
\end{equation}
where we have used \eqref{Z41}, \eqref{z44} and \eqref{Y55}. Thus, after quick computation using \eqref{constraint}, \eqref{x42} and \eqref{Y55}, the metric \eqref{ansmet} can be cast into 
\begin{equation}\label{ansmetbaru}
 ds^2=-N^2 dt^2 + \frac{dR^2}{\left(\frac{1}{2\kappa_D^2\xi} +\frac{C}{R^{D-4}}\right) R^2Y} + R^2d\Omega^2_{(D-2)} ~ .
\end{equation}
Then, the Ricci tensor \eqref{compRiccitens} are simplified into
\begin{eqnarray}\label{compRiccitensbr}
		{{\mathcal{R}}_{00}}&&=-{{g}_{00}}\left\{ \frac{1}{4}\left( \frac{1}{2\kappa _{D}^{2}\xi }+\frac{C}{{{e}^{\left( D-4 \right)x}}} \right)\left( \zeta \sqrt{\frac{Y}{Z}}\frac{dZ}{dx}-\frac{dY}{dx}+Y+\zeta Z-2\sqrt{\zeta YZ} \right) \right. \notag\\ 
		&& \left. -\frac{C\left( D-4 \right)}{4{{e}^{\left( D-4 \right)x}}}\left( \sqrt{\zeta YZ}-Y \right)+\frac{\left( D-2 \right)}{2}\left( \frac{1}{2\kappa _{D}^{2}\xi }+\frac{C}{{{e}^{\left( D-4 \right)x}}} \right)\left( \sqrt{\zeta YZ}-Y \right) \right\} ~ , \notag\\
		&&\notag\\
		{{\mathcal{R}}_{11}}&&=-{{g}_{11}}\left\{ \frac{1}{4}\left( \frac{1}{2\kappa _{D}^{2}\xi }+\frac{C}{{{e}^{\left( D-4 \right)x}}} \right)\left( \zeta \sqrt{\frac{Y}{Z}}\frac{dZ}{dx}-\frac{dY}{dx}+Y+\zeta Z-2\sqrt{\zeta YZ} \right) \right. \notag\\ 
		&&-\frac{C\left( D-4 \right)}{4{{e}^{\left( D-4 \right)x}}}\left( \sqrt{\zeta YZ}-Y \right)+(D-2)\left( \frac{1}{2\kappa _{D}^{2}\xi }+\frac{C}{{{e}^{\left( D-4 \right)x}}} \right)\left( {{Y}^{1/2}}\frac{dY}{dx}+Y \right)\notag\\
		&&\left. +\frac{C(D-2)\left( D-4 \right)}{2}Y  \right\} ~ , \notag\\
		{{\mathcal{R}}_{ij}}&&={{{\hat{\mathcal{R}}}}_{ij}} - g_{ij}\left\{ \left( {{Y}^{1/2}}\frac{dY}{dx}+Y \right)\left( \frac{1}{2\kappa _{D}^{2}\xi }+\frac{C}{{{e}^{\left( D-4 \right)x}}} \right)-\frac{C\left( D-4 \right)}{2{{e}^{\left( D-4 \right)x}}}Y \right. \notag\\ 
		&&+\left( \sqrt{\zeta YZ}-Y \right)\left( \frac{1}{2\kappa _{D}^{2}\xi }+\frac{C}{{{e}^{\left( D-4 \right)x}}} \right)\left. -Y\left( \frac{1}{2\kappa _{D}^{2}\xi }+\frac{C}{{{e}^{\left( D-4 \right)x}}} \right)\left( D-3 \right) \right\} ~ , 	  
\end{eqnarray}
where ${{g}_{00}} \equiv  -N^2$, ${{g}_{11}} \equiv \frac{e^{-2x}}{\left(\frac{1}{2\kappa_D^2\xi} +\frac{C}{e^{(D-4)x}}\right) Y}$, and $g_{ij} \equiv e^{2x} \hat{g}_{ij}$. The Ricci scalar curvature \eqref{Riccisclr1}, and the Kretschman scalar \eqref{NormRiem} can be recast into
\begin{eqnarray} \label{Riccisclr1br}
\mathcal{R}&=& e^{-2x}  \hat{\mathcal{R}}  + \frac{1}{4}\left( \frac{1}{2\kappa _{D}^{2}\xi }+\frac{C}{{{e}^{\left( D-4 \right)x}}} \right)\left( \zeta \sqrt{\frac{Y}{Z}}\frac{dZ}{dx}-\frac{dY}{dx}+Y+\zeta Z-2\sqrt{\zeta YZ} \right) \notag\\ 
	  	&-&\frac{C\left( D-4 \right)}{4{{e}^{\left( D-4 \right)x}}}\left( \sqrt{\zeta YZ}-Y \right)+\frac{\left( D-2 \right)}{2}\left( \frac{1}{2\kappa _{D}^{2}\xi }+\frac{C}{{{e}^{\left( D-4 \right)x}}} \right)\left( \sqrt{\zeta YZ}+{{Y}^{1/2}}\frac{dY}{dx} \right) \notag\\ 
	  	&+&\frac{C(D-2)\left( D-4 \right)}{2}Y-Y\left( \frac{1}{2\kappa _{D}^{2}\xi }+\frac{C}{{{e}^{\left( D-4 \right)x}}} \right)\left( D-3 \right)  ~ ,  
\end{eqnarray}
\begin{eqnarray}\label{NormRiembr}
		\|Riem{{\|}^{2}}&=&4\left\{ \frac{1}{4}\left( \frac{1}{2\kappa _{D}^{2}\xi }+\frac{C}{{{e}^{\left( D-4 \right)x}}} \right)\left( \zeta \sqrt{\frac{Y}{Z}}\frac{dZ}{dx}-\frac{dY}{dx}+Y+\zeta Z-2\sqrt{\zeta YZ} \right) \right. \notag\\ 
		&-&{{\left. \frac{\left( D-4 \right)C}{4}\left( \sqrt{\zeta YZ}-Y \right) \right\}}^{2}}+4(D-2)\left\{ \left( {{Y}^{1/2}}\frac{dY}{dx}+Y \right)\left( \frac{1}{2\kappa _{D}^{2}\xi }+\frac{C}{{{e}^{\left( D-4 \right)x}}} \right) \right. \notag\\ 
		&-&{{\left. \frac{\left( D-4 \right)C}{2}Y-\left( D-3 \right)\left( \frac{1}{2\kappa _{D}^{2}\xi }+\frac{C}{{{e}^{\left( D-4 \right)x}}} \right)Y \right\}}^{2}} \notag\\ 
		&+&(D-2){{\left( \sqrt{\zeta Z}-{{Y}^{1/2}} \right)}^{2}}Y{{\left( \frac{1}{2\kappa _{D}^{2}\xi }+\frac{C}{{{e}^{\left( D-4 \right)x}}} \right)}^{2}}+ e^{-4x} {{\widehat{\mathcal{R}}}^{ijkl}}{{\widehat{\mathcal{R}}}_{ijkl}}  ~ ,  
\end{eqnarray}
respectively. These  will be useful for our next discussion.

At the end, some comments are in order. First, the $(D-2)$-dimensional submanifold ${\mathcal S}^{D-2}$  belongs to a special class in which it admits constant triplet structure $\left( \hat{T}, \frac{ {\mathcal{E}^{i}}_{,i}}{\sqrt{\hat{g}}}, \frac{{{\hat{e}}^{\bar{b}}}_{i}\left( {\mathcal{F}^{ik}_{\bar{b}}}\right) _{,k}}{\sqrt{\hat{g}}} \right)$ given in \eqref{torsionconstants}. Some common examples do exist such as 2-sphere $S^2$, $(D-2)$-dimensional torus $T^{D-2}$, and $\lR^{D-2}$ together with their product spaces. For $S^2$, we have $(0,-1,0)$ and the remaining two example have $(0,0,0)$. These examples are spaces of constant scalar curvature. Second, since the master equation \eqref{mastereq} is extremely complicated to solve, we can generally study its behavior in some regions. If the solutions were physical black holes that admit cosmic censorship conjecture, then these regions are the horizon and the outer region such as asymptotic region. Otherwise the spacetimes may have naked singularity at finite $r$ or be smooth everywhere.

 \section{Special Class of Exact Solutions}
\label{sec:specsolexact}
 In this section we discuss a special class of exact solutions of the master equation \eqref{mastereq}. The first part is to focus on four dimensional model, while in the second part we consider higher dimensional model.
 
 \subsection{Four Dimensional Case}
 In four dimensions, we only consider a case where the scalar potential has the form of the sinh-Gordon potential
 \begin{equation}\label{potential4d}
 	V\left( x \right)=\alpha {{e}^{-2x}}+\gamma ~ ,
 \end{equation}
 where $\alpha, \gamma$ are constants. In this case, by considering (\ref{toreq4}),  the constraint of the theory has to be of the form
 \begin{equation}\label{constraintd4}
 	\phi'=\frac{\nu}{K} ~ ,
 \end{equation}
 with $\nu$ is a constant. Then, we insert (\ref{constraintd4}) to (\ref{toreq8})-(\ref{toreq10}), the equation (\ref{toreq8})-(\ref{toreq10}) can be cast into the following form:
 \begin{equation}\label{d4eom1}
 	\frac{dY}{dx}+3Y-\frac{{{e}^{-2x}}}{{{\nu }^{2}}}+\frac{\kappa _{4}^{2}\left( {{\nu }^{2}}+2V \right)}{2{{\nu }^{2}}\left( 1-2\xi {{\nu }^{2}}\kappa _{4}^{2} \right)}=0 ~ ,
 \end{equation}
 \begin{equation}\label{d4eom2}
 	 \sqrt{\zeta YZ}-\frac{\left( 2\xi {{\nu }^{2}}\kappa _{4}^{2}-1 \right){{e}^{-2x}}}{{{\nu }^{2}}\left( 6\xi {{\nu }^{2}}\kappa _{4}^{2}-1 \right)}+\frac{\kappa _{4}^{2}\left( {{\nu }^{2}}-2V \right)}{2{{\nu }^{2}}\left( 6\xi {{\nu }^{2}}\kappa _{4}^{2}-1 \right)}=0 ~,
 \end{equation}
 \begin{equation}\label{d4eom3}
 	\zeta \sqrt{\frac{Y}{Z}}\frac{dZ}{dx}+{{\zeta}}Z+\frac{{{e}^{-2x}}}{{{\nu }^{2}}}+\frac{3\kappa _{4}^{2}\left( {{\nu }^{2}}+2V \right)}{2{{\nu }^{2}}\left( 1-2\xi {{\nu }^{2}}\kappa _{4}^{2} \right)}=0 ~.
 \end{equation}
 Making use (\ref{potential4d}), we have the solution of (\ref{d4eom1})  
 \begin{equation}\label{sold4Y}
 	Y\left( x \right)=-{{a}_{1}}{{e}^{-2x}}-\frac{{{a}_{2}}}{3}+{{a}_{3}}{{e}^{-3x}} ~ ,
 \end{equation}
 where  
 \begin{eqnarray}
 	&& {{a}_{1}}=-\frac{1}{{{\nu }^{2}}}+\frac{\kappa _{4}^{2}\alpha }{{{\nu }^{2}}\left( 1-2\xi {{\nu }^{2}}\kappa _{4}^{2} \right)}~, \\ 
 	&& {{a}_{2}}=\frac{\kappa _{4}^{2}\left( {{\nu }^{2}}+2\gamma  \right)}{2{{\nu }^{2}}\left( 1-2\xi {{\nu }^{2}}\kappa _{4}^{2} \right)} ~,
 \end{eqnarray}
and $a_3$ is the integration constant. 

 Next, using the result in (\ref{sold4Y}) and taking simply   $a_3=0$, and then inserting to (\ref{d4eom2}), we obtain
 \begin{equation}\label{sold4Z}
 	Z(x)=-\frac{{{\left( {{b}_{1}}{{e}^{-2x}}+{{b}_{2}} \right)}^{2}}}{\zeta \left( {{a}_{1}}{{e}^{-2x}}+\frac{{{a}_{2}}}{3}\right)}  ~ ,
 \end{equation}
 with
 \begin{eqnarray}
 	&& {{b}_{1}}=\frac{2\xi {{\nu }^{2}}\kappa _{4}^{2}-1+\kappa _{4}^{2}\alpha }{{{\nu }^{2}}\left( 6\xi {{\nu }^{2}}\kappa _{4}^{2}-1 \right)} ~,\\ 
 	&& {{b}_{2}}=\frac{\kappa _{4}^{2}\left( 2\gamma -{{\nu }^{2}} \right)}{2{{\nu }^{2}}\left( 6\xi {{\nu }^{2}}\kappa _{4}^{2}-1 \right)} ~.	
 \end{eqnarray}
  By inserting (\ref{sold4Y}) and (\ref{sold4Z}) to (\ref{d4eom3}), we find the following equations 
  \begin{eqnarray}\label{syaratzeta}
 	2{{\zeta }^{1/2}}{{b}_{1}}{{a}_{1}}+b_{1}^{2}+{{a}_{1}}{{c}_{1}}=0  ~,
 \end{eqnarray} 
 \begin{equation}\label{eomalpha}
 	\frac{4}{3}{{\zeta }^{1/2}}{{b}_{1}}{{a}_{2}}-2{{\zeta }^{1/2}}{{a}_{1}}{{b}_{2}}+2{{b}_{1}}{{b}_{2}}+{{a}_{1}}{{c}_{2}}+\frac{1}{3}{{a}_{2}}{{c}_{1}}=0  ~, 
 \end{equation}
 \begin{equation}\label{eomgamma}
 	b_{2}^{2}+\frac{1}{3}{{a}_{2}}{{c}_{2}}=0   ~,	
 \end{equation}
where
\begin{eqnarray}
	&& {{c}_{1}}=-\left( \frac{1}{{{\nu }^{2}}}+\frac{3\kappa _{4}^{2}\alpha }{{{\nu }^{2}}\left( 1-2\xi {{\nu }^{2}}\kappa _{4}^{2} \right)} \right) ~, \\ 
	&& {{c}_{2}}=-\frac{\left( 3\kappa _{4}^{2}{{\nu }^{2}}+6\kappa _{4}^{2}\gamma  \right)}{2{{\nu }^{2}}\left( 1-2\xi {{\nu }^{2}}\kappa _{4}^{2} \right)} ~ , 	
\end{eqnarray}
such that the constant $\gamma$ is given by  
 \begin{equation}\label{gamma}
 	\gamma =\frac{-\left( 4+\frac{{{{\tilde{\eta_4 }}}^{2}}}{{{\eta_4 }^{2}}} \right)\pm \sqrt{{{\left( 4+\frac{{{{\tilde{\eta_4 }}}^{2}}}{{{\eta_4 }^{2}}} \right)}^{2}}-4{{\left( 4-\frac{{{{\tilde{\eta_4 }}}^{2}}}{{{\eta_4 }^{2}}} \right)}^{2}}}}{2\left( 4-\frac{{{{\tilde{\eta_4 }}}^{2}}}{{{\eta_4 }^{2}}} \right)}  ~ ,
 \end{equation}
 where
 \begin{eqnarray}
 	& \eta_4 =\left( 1-2\xi {{\nu }^{2}}\kappa _{4}^{2} \right)  ~ , \\ 
 	& \tilde{\eta_4 }=\left( 6\xi {{\nu }^{2}}\kappa _{4}^{2}-1 \right)	 ~ .
 \end{eqnarray}
For (\ref{gamma}) to be real constant,  the constant $\nu$ has to satisfy 
\begin{equation}
	\nu \ge \sqrt{\frac{3+2\sqrt{3}}{\left( 18+4\sqrt{3} \right)\xi \kappa _{4}^{2}}}  ~ ,
\end{equation}
while from equation (\ref{syaratzeta}), one can obtain the conditions for $\zeta$
 \begin{equation}\label{zeta}
 \zeta ={{\left( \frac{{{b}_{1}}}{2{{a}_{1}}}+\frac{{{c}_{1}}}{2{{b}_{1}}} \right)}^{2}}   ~,
 \end{equation}
such that from (\ref{eomalpha}) and  (\ref{zeta}), we get 
\begin{eqnarray}
	\alpha &=&{{\left( -\frac{2d_{2}^{3}}{54d_{1}^{3}}+\frac{{{d}_{2}}{{d}_{3}}}{6d_{1}^{2}}-\frac{{{d}_{4}}}{2{{d}_{1}}}+\sqrt{{{\left( -\frac{2d_{2}^{3}}{54d_{1}^{3}}+\frac{{{d}_{2}}{{d}_{3}}}{6d_{1}^{2}}-\frac{{{d}_{4}}}{2{{d}_{1}}} \right)}^{2}}+{{\left( -\frac{d_{2}^{2}}{9d_{1}^{2}}+\frac{{{d}_{3}}}{3{{d}_{1}}} \right)}^{3}}} \right)}^{1/3}} \notag\\ 
	&& -\left( \frac{2d_{2}^{3}}{54d_{1}^{3}}-\frac{{{d}_{2}}{{d}_{3}}}{6d_{1}^{2}}+\frac{{{d}_{4}}}{2{{d}_{1}}}+\sqrt{{{\left( -\frac{2d_{2}^{3}}{54d_{1}^{3}}+\frac{{{d}_{2}}{{d}_{3}}}{6d_{1}^{2}}-\frac{{{d}_{4}}}{2{{d}_{1}}} \right)}^{2}}+{{\left( -\frac{d_{2}^{2}}{9d_{1}^{2}}+\frac{{{d}_{3}}}{3{{d}_{1}}} \right)}^{3}}} \right)\notag  ~ .\\
\end{eqnarray}
where
\begin{eqnarray}
		&& {{d}_{1}}=\frac{2\kappa _{4}^{6}{{a}_{2}}}{3{{{\tilde{\eta_4 }}}^{3}}}-\frac{3\kappa _{4}^{6}{{a}_{2}}}{\tilde{\eta_4 }{{\eta_4 }^{2}}}+\frac{3\kappa _{4}^{6}{{b}_{2}}}{{{{\tilde{\eta_4 }}}^{2}}\eta_4 }+\frac{3\kappa _{4}^{6}{{b}_{2}}}{{{\eta_4 }^{3}}}+\frac{\kappa _{4}^{6}{{c}_{2}}}{\tilde{\eta_4 }{{\eta_4 }^{2}}}  ~, \\ 
		&& {{d}_{2}}=-\frac{6\kappa _{4}^{4}{{a}_{2}}\eta_4 }{3{{{\tilde{\eta_4 }}}^{3}}}+\frac{5\kappa _{4}^{4}{{a}_{2}}}{\tilde{\eta_4 }\eta_4 }-\frac{9\kappa _{4}^{4}{{b}_{2}}}{{{{\tilde{\eta_4 }}}^{2}}}-\frac{5\kappa _{4}^{4}{{b}_{2}}}{{{\eta_4 }^{2}}}-\frac{3\kappa _{4}^{4}{{c}_{2}}}{\tilde{\eta_4 }\eta_4 }  ~, \\ 
		&& {{d}_{3}}=\frac{6\kappa _{4}^{2}{{a}_{2}}{{\eta_4 }^{2}}}{3{{{\tilde{\eta_4 }}}^{3}}}-\frac{\kappa _{4}^{2}{{a}_{2}}}{{\tilde{\eta_4 }}}+\frac{9\kappa _{4}^{2}{{b}_{2}}\eta_4}{{{{\tilde{\eta_4 }}}^{2}}}+\frac{\kappa _{4}^{2}{{b}_{2}}}{\eta_4 }+\frac{3\kappa _{4}^{2}{{c}_{2}}}{{\tilde{\eta_4 }}}  ~, \\ 
		&& {{d}_{4}}=-\frac{{{\eta_4 }^{3}}}{3{{{\tilde{\eta }}}^{3}}}-\frac{{{a}_{2}}\eta_4 }{{\tilde{\eta_4 }}}-\frac{3{{\eta_4 }^{2}}{{b}_{2}}}{{{{\tilde{\eta_4 }}}^{2}}}+{{b}_{2}}-\frac{{{c}_{2}}\eta)_4 }{{\tilde{\eta_4}}}  ~.
\end{eqnarray}

As we see from \eqref{sold4Y}, this model admits singularity at the origin (as $x \to -\infty$). Moreover, the Kretschman scalar \eqref{NormRiembr} diverges at this point. On the other hand, in the asymptotic region (as $x \to +\infty$) the spacetime converges to the space of constant scalar curvature but not Einstein.

To cast  the scalar field $\phi$ as the function of $x$, we use (\ref{Y55}) and (\ref{sold4Y}) so that
\begin{equation}\label{phi}
	\phi (x)=-\frac{\sqrt{3}\text{ }}{\sqrt{-{{a}_{2}}}} \tan^{-1} \left( \frac{\sqrt{3{{a}_{1}}{{e}^{-2x}}+{{a}_{2}}}}{\sqrt{-{{a}_{2}}}} \right)+c ~ .
\end{equation}
Then, the potential in (\ref{potential4d}) can also be expressed in terms of its scalar field
\begin{equation}
V\left( \phi  \right)=-\frac{\kappa _{4}^{2}\left( {{\nu }^{2}}+2\gamma  \right)\alpha }{6\left( 2\xi {{\nu }^{2}}\kappa _{4}^{2}-1+\kappa _{4}^{2}\alpha  \right)}\left[ {{\tan }^{2}}\left( -\frac{{{\kappa }_{4}}{{\left( {{\nu }^{2}}+2\gamma  \right)}^{1/2}}}{\sqrt{6}\nu {{\left( 2\xi {{\nu }^{2}}\kappa _{4}^{2}-1 \right)}^{1/2}}}\phi  \right)+1 \right]+\gamma ~.
\end{equation}

 \subsection{Higher Dimensional Case}
 For higher dimensional model, we particularly consider a case where  $C=0$ in (\ref{constraint}). Then, from (\ref{toreq1}), we find that
 \begin{equation}
 	V=-\frac{1}{4\kappa _{D}^{2}\xi } ~ ,
 \end{equation}
while from (\ref{toreq2}) and (\ref{toreq3}) we have
\begin{equation}\label{y1}
	Y+\frac{1}{(D-4)}\sqrt{\zeta YZ}-\frac{2\kappa _{D}^{2}\xi {{\lambda }_{1}}}{(D-2)(D-4){{e}^{2x}}}+\frac{\kappa _{D}^{2}}{(D-2)(D-4)}=0 ~ ,
\end{equation}
\begin{eqnarray}\label{y2}
	&&Y+\frac{1}{(D-4)}\sqrt{\zeta YZ}-\frac{\sqrt{2}{{\lambda }_{2}}{{\xi }^{1/2}}\kappa _{D}^{3/2}}{\left( D-4 \right)\left( D-2 \right)N{{e}^{\left( D-2 \right)x}}}+\frac{\kappa _{D}^{2}}{\left( D-4 \right)\left( D-2 \right)}-\frac{2\kappa _{D}^{2}\xi {{\lambda }_{1}}}{\left( D-4 \right)}{{e}^{-2x}}=0 ~ ,\notag\\
	&&
\end{eqnarray}
which imply
\begin{equation}\label{N}
	N(x)=-\frac{{{\lambda }_{2}}}{\sqrt{2{{\kappa }_{D}}\xi }{{\lambda }_{1}}\left( D-3 \right)\left(D-2\right){{e}^{\left( D-4 \right)x}}} ~ .
\end{equation}
Using  (\ref{z44}) and  (\ref{N}), it follows
\begin{equation}\label{Z}
	Z=\frac{{{\left( 2D-9 \right)}^{2}}Y}{\zeta } ~ .
\end{equation}
Finally, inserting (\ref{Z}) to (\ref{y1}) we then have
\begin{equation}\label{Y}
Y\left( x \right)=-\frac{2\kappa _{D}^{2}\xi {{\lambda }_{1}}}{(D-2)(D-5){{e}^{2x}}}+\frac{\kappa _{D}^{2}}{(D-2)(D-5)} ~ ,
\end{equation}
showing that $Y$ converges to a constant  for $x\to +\infty$. However, $Y$ diverges at the origin (as  $x\to -\infty$). These show that the model has  singularity at the origin since the Kretschman scalar \eqref{NormRiembr} blows up, while in the asymptotic region the spacetime becomes the space of constant scalar curvature but not Einstein.

For the case of $C>0$, it is however difficult to find an exact solution due to  highly nonlinearity of the master equation \eqref{mastereq}. Therefore, we only consider asymptotic solutions at large distances and construct the global existence of \eqref{mastereq} in the next sections.

 \section{Properties of Static Spacetimes}
 \label{sec:propertiesstatic}

This section is divided into two parts. In the first part, we prove that there is no black hole solution in the model by analyzing the Kretschmann scalar. The second part contains the discussion of the behavior of solutions in the asymptotic region.
 \subsection{No Horizon Limit}
 \label{subsec:nohorlim}

First of all, let us write down the spatial part of the metric \eqref{ansmetbaru}
\begin{equation}\label{ansmetbaruspa}
 ds^2 |_{t = const.}=\frac{dR^2}{\left(\frac{1}{2\kappa_D^2\xi} +\frac{C}{R^{D-4}}\right) R^2Y}+R^2d\Omega^2_{(D-2)} ~ ,
\end{equation}
defined on the submanifold $\Sigma^{D-1}$. As ${\mathcal S}^{D-2}$ can also be viewed as the submanifold of $\Sigma^{D-1}$, the associated mean curvature $H$ has the form
\begin{equation} \label{meancurv}
 H =  \left(\frac{1}{2\kappa_D^2\xi} +\frac{C}{R^{D-4}}\right)^{1/2}Y^{1/2} ~,
\end{equation}
with $H \ge 0$. If the spacetime ${\mathcal M}^D$ is a black hole, then there exists a region called (event) horizon with radius $r_H > 0$ where ${\mathcal S}^{D-2}$ is a minimal submanifold, \textit{i.e.} $H = 0$, see for example \cite{Chrusciel:2010fn}, which implies 
\begin{equation} \label{meancurv=0}
  Y = 0 ~ ,
\end{equation}
since $\xi > 0$ and $C \ge 0$. At the horizon,   ${\mathcal M}^D$ breaks into ${\mathcal N}^2 \times {\mathcal S}^{D-2}$ where ${\mathcal N}^2$ is a two dimensional surface which could be either anti-de Sitter (AdS) surface or flat Minkowski surface \cite{Kunduri:2007}. This implies that the invariant quantity such as the Kretschmann scalar \eqref{NormRiem} and the Ricci scalar \eqref{Riccisclr1} must be finite at this region. It is worth noticing that the quantity $H$ is defined on the normal bundle on ${\mathcal S}^{D-2}$ which is different to the Riemann curvature, the Ricci (curvature) tensor, and the Ricci scalar curvature defined on the spacetime ${\mathcal M}^D$ and the submanifold ${\mathcal S}^{D-2}$.
\begin{theorem}
	\label{thmmainresnohor}
Suppose we have a static spacetime ${\mathcal M}^D$ with metric \eqref{ansmetbaru} satisfying the master equation  \eqref{mastereq} with $\xi > 0$ and $C \ge 0$. Then, the submanifold ${\mathcal S}^{D-2} \subseteq {\mathcal M}^D$ is not minimal which implies that ${\mathcal M}^D$ is not a black hole.
\end{theorem}
\begin{proof}
Let us first consider the Kretschmann scalar \eqref{NormRiembr}. Since our metric is static given by \eqref{ansmetbaru}, one can make a coordinate transformation with respect to $R(r) = e^x$ such that we have $R(r) = r$.  Suppose there exists a horizon with radius $r_H > 0$ such that near the horizon $Y=Y(r_H)+Y'(r_H)(r-r_H)$ where $Y(r=r_H)=0$. We can define, $\varepsilon \equiv r-r_H$ and $Y'(r_H)=C$ where $C$ is a constant so that  $Y(r)=C\varepsilon(r)$. From \eqref{ZeqY}, we find $Z \propto 1/ \varepsilon$ near the region. Then, we have $N'/N \propto 1/ \varepsilon$ and $N''/N \propto 1/ \varepsilon^2 $. Inserting all the data to \eqref{NormRiembr}, we obtain $\| Riem\|^2 \propto 1/ \varepsilon$. Evaluating at $r = r_H$ means that we take the limit $\varepsilon \to 0$, which implies that \eqref{NormRiembr} diverges. Thus, the submanifold ${\mathcal S}^{D-2}$ is not minimal. 
\end{proof}
We want to write some comments as follows. Using the above arguments, one could show that at $r=r_H$ with $Y=0$ the Ricci scalar also diverges. In other words, any point besides the origin ($r = r_a > 0$) with $Y = 0$ is naked singularity on ${\mathcal M}^D$. Second, at $Y=0$ \eqref{toreq10} becomes ill defined because its left hand side blows up. In view of \eqref{Y55}, this point is related to $\phi' \to + \infty$ because $K(r) \to 0$ (see also \eqref{constraint}). Thirdly, it is easy to see from \eqref{NormRiembr} that if $Y \to c \ge 0$ at the origin ($r = 0$), then the origin is also naked singularity on ${\mathcal M}^D$.

Finally, let us discuss the $C< 0$ case with $D > 4$. Looking at \eqref{meancurv}, it might be a point $r = \tilde{r}_H > 0$ such that $ R^{D-4}(\tilde{r}_H) = -2\kappa_D^2\xi C $. From \eqref{constraint}) and \eqref{Y55}, at this point one would have $\phi' \to 0$ implying $Y \to + \infty$, unless $R'(\tilde{r}_H) = 0$ which makes $Y$ undetermined. However, such a condition contradicts with the property of the static metric \eqref{ansmetbaru} that one can have $R(r) = r$. Therefore, the case of $C< 0$ and $D > 4$ should be excluded.

In conclusion, we can write
\begin{corollary}
	\label{cornakedsing}
The static spacetime $ {\mathcal M}^D$ admits at least a naked singularity at the origin.
\end{corollary}
%

%


 \normalsize

\subsection{Solutions at Large Distances}
In this section, we want to solve perturbatively the equation \eqref{mastereq} by expanding $Y(x)=Y_0+Y_1(x)$ with $|Y_0| \gg |Y_1|$ as $x \to +\infty$. This has been studied in \cite{kofinas2012torsi, Yaqin2017comment} for $D=4$ case in which we could have $|Y_0| \gg |Y_1|$, for example, by taking large $\xi$ or large $C$ as discussed above. 
\subsubsection{Asymptotic Solution in Four Dimensions}
Let us first discuss the  $D = 4$ case. In this case, we split the master equation \eqref{mastereq} into the zero and the first order as follows
\begin{eqnarray}\label{eqY0d4}
			&& \left\{ \frac{18\kappa _{4}^{2}{{\xi }^{2}}{{C}^{2}}}{2{{{\tilde{\eta }}}_{4}}}-\frac{18{{{\tilde{\eta }}}_{4}}}{\kappa _{4}^{2}} \right\}\frac{{{Y}_{0}}^{2}}{{{\zeta }^{1/2}}}+\frac{4\kappa _{4}^{2}}{{{\zeta }^{1/2}}}{{{\tilde{\eta }}}_{4}}{{\left( -\left( \frac{{{\lambda }_{1}}}{2}+{{\lambda }_{2}} \right)\frac{\kappa _{4}^{2}\xi }{2{{{\tilde{\eta }}}_{4}}{{\eta }_{4}}}-\frac{1}{2{{{\tilde{\eta }}}_{4}}{{\eta }_{4}}}+\frac{\kappa _{4}^{2}\xi C}{{{{\tilde{\eta }}}_{4}}{{\eta }_{4}}} \right)}^{2}} \notag\\ 
			&& -\left\{ \frac{12\kappa _{4}^{2}\xi C}{{{\zeta }^{1/2}}}\left( -\left( \frac{{{\lambda }_{1}}}{2}+{{\lambda }_{2}} \right)\frac{\kappa _{4}^{2}\xi }{2{{{\tilde{\eta }}}_{4}}{{\eta }_{4}}}-\frac{1}{2{{{\tilde{\eta }}}_{4}}{{\eta }_{4}}}+\frac{\kappa _{4}^{2}\xi C}{{{{\tilde{\eta }}}_{4}}{{\eta }_{4}}} \right)+\frac{6}{C}\left( \frac{{{\lambda }_{1}}}{2}+{{\lambda }_{2}} \right)\frac{{{{\tilde{\eta }}}_{4}}}{{{\eta }_{4}}{{\zeta }^{1/2}}} \right\}{{Y}_{0}}=0~,\notag\\
			&&	
\end{eqnarray}
\begin{eqnarray}\label{eqY1d4}
			&& \frac{d^2 Y_1}{dx^2}  + \frac{d Y_1}{dx} \left\{ -\frac{4\kappa _{4}^{2}\xi C}{{{\eta }_{4}}} \right.-12\xi C+\frac{2}{{{\eta }_{4}}}+\frac{2\kappa _{4}^{2}\xi }{{{\eta }_{4}}}\left( \frac{{{\lambda }_{1}}}{2}+{{\lambda }_{2}} \right)-6\xi C{{Y}_{0}}-\frac{{{{\tilde{\eta }}}_{4}}}{2\kappa _{4}^{2}C\xi {{\zeta }^{1/2}}}{{Y}_{0}} \notag\\ 
			&& \left. +\frac{1}{{{\zeta }^{1/2}}}\left( -\left( \frac{{{\lambda }_{1}}}{2}+{{\lambda }_{2}} \right)\frac{\kappa _{4}^{4}\xi }{2{{{\tilde{\eta }}}_{4}}{{\eta }_{4}}}-\frac{\kappa _{4}^{2}}{2{{{\tilde{\eta }}}_{4}}{{\eta }_{4}}}+\frac{\kappa _{4}^{4}\xi C}{{{{\tilde{\eta }}}_{4}}{{\eta }_{4}}} \right)+\frac{3\kappa _{4}^{2}\xi C}{2{{{\tilde{\eta }}}_{4}}{{\zeta }^{1/2}}}{{Y}_{0}} \right\}+{{Y}_{1}}\left\{ \frac{9\kappa _{4}^{2}{{\xi }^{2}}{{C}^{2}}}{{{{\tilde{\eta }}}_{4}}{{\zeta }^{1/2}}}{{Y}_{0}}-\frac{36{{{\tilde{\eta }}}_{4}}}{{{\zeta }^{1/2}}\kappa _{4}^{2}}{{Y}_{0}} \right. \notag\\ 
			&& \left. -\frac{6{{{\tilde{\eta }}}_{4}}}{{{\eta }_{4}}{{\zeta }^{1/2}}C}\left( \frac{{{\lambda }_{1}}}{2}+{{\lambda }_{2}} \right)-\frac{12\xi C}{{{\zeta }^{1/2}}}\left( -\left( \frac{{{\lambda }_{1}}}{2}+{{\lambda }_{2}} \right)\frac{\kappa _{4}^{4}\xi }{2{{{\tilde{\eta }}}_{4}}{{\eta }_{4}}}-\frac{\kappa _{4}^{2}}{2{{{\tilde{\eta }}}_{4}}{{\eta }_{4}}}+\frac{\kappa _{4}^{4}\xi C}{{{{\tilde{\eta }}}_{4}}{{\eta }_{4}}} \right) \right\}=0~,
\end{eqnarray}
respectively. From (\ref{eqY0d4}) we get
\begin{equation}\label{solY0d4}
{{Y}_{0}}=\frac{4\kappa _{4}^{2}{{\eta }_{4}}{{{\tilde{\eta }}}_{4}}{{\left( -\left( \frac{{{\lambda }_{1}}}{2}+{{\lambda }_{2}} \right)\frac{\kappa _{4}^{2}\xi }{2{{{\tilde{\eta }}}_{4}}{{\eta }_{4}}}-\frac{1}{2{{{\tilde{\eta }}}_{4}}{{\eta }_{4}}}+\frac{\sqrt{2}}{{{\eta }_{4}}} \right)}^{2}}}{12\sqrt{2}\left( -\left( \frac{{{\lambda }_{1}}}{2}+{{\lambda }_{2}} \right)\frac{\kappa _{4}^{2}\xi }{2}-\frac{1}{2}+\sqrt{2} \right)+\frac{6\kappa _{4}^{2}\xi }{\sqrt{2}}\left( \frac{{{\lambda }_{1}}}{2}+{{\lambda }_{2}} \right)}~.
\end{equation}
As can be seen from the preceding discussion, $Y_0>0$, so in order for this condition to be met, the constants $C$ and $\left(\frac{{{\lambda }_{1}}}{2}+{{\lambda }_{2}}\right)$ must satisfy the following condition
\begin{eqnarray}\label{Cdanlambdad4asimtotik}
		C&=&\frac{\sqrt{2}{{{\tilde{\eta }}}_{4}}}{\kappa _{4}^{2}\xi }~,\\
		\left( \frac{{{\lambda }_{1}}}{2}+{{\lambda }_{2}} \right)&<&\frac{2-4\sqrt{2}}{\kappa _{4}^{2}\xi }~.	
\end{eqnarray}
The spacetime at the lowest order has constant Ricci scalar curvature but not Einstein.

For the first order, we obtain the solution shown below
\begin{equation}
	Y_1(x)=C_1 e^{-\alpha x} ~ ,
\end{equation}
with $\alpha$ being a real positive constant that fulfils  the following condition
\begin{equation}\label{alpha}
	\alpha =\frac{{{m}_{1}}+\epsilon \sqrt{{{m}_{1}}^{2}-4{{m}_{2}}}}{2}
\end{equation}
where $\epsilon=\pm1$ and
\begin{eqnarray}
			{{m}_{1}}&=&-\frac{4\kappa _{4}^{2}\xi C}{{{\eta }_{4}}}-12\xi C+\frac{2}{{{\eta }_{4}}}+\frac{2\kappa _{4}^{2}\xi }{{{\eta }_{4}}}\left( \frac{{{\lambda }_{1}}}{2}+{{\lambda }_{2}} \right)-6\xi C{{Y}_{0}}-\frac{{{{\tilde{\eta }}}_{4}}}{2\kappa _{4}^{2}C\xi {{\zeta }^{1/2}}}{{Y}_{0}} \notag\\ 
			&&+\frac{1}{{{\zeta }^{1/2}}}\left( -\left( \frac{{{\lambda }_{1}}}{2}+{{\lambda }_{2}} \right)\frac{\kappa _{4}^{4}\xi }{2{{{\tilde{\eta }}}_{4}}{{\eta }_{4}}}-\frac{\kappa _{4}^{2}}{2{{{\tilde{\eta }}}_{4}}{{\eta }_{4}}}+\frac{\kappa _{4}^{4}\xi C}{{{{\tilde{\eta }}}_{4}}{{\eta }_{4}}} \right)+\frac{3\kappa _{4}^{2}\xi C}{2{{{\tilde{\eta }}}_{4}}{{\zeta }^{1/2}}}{{Y}_{0}}~, \\ 
			&&\notag\\
			{{m}_{2}}&=&-\frac{6{{{\tilde{\eta }}}_{4}}}{{{\eta }_{4}}{{\zeta }^{1/2}}C}\left( \frac{{{\lambda }_{1}}}{2}+{{\lambda }_{2}} \right)-\frac{12\xi C}{{{\zeta }^{1/2}}}\left( -\left( \frac{{{\lambda }_{1}}}{2}+{{\lambda }_{2}} \right)\frac{\kappa _{4}^{4}\xi }{2{{{\tilde{\eta }}}_{4}}{{\eta }_{4}}}-\frac{\kappa _{4}^{2}}{2{{{\tilde{\eta }}}_{4}}{{\eta }_{4}}}+\frac{\kappa _{4}^{4}\xi C}{{{{\tilde{\eta }}}_{4}}{{\eta }_{4}}} \right) \notag\\
			&&-\frac{9\kappa _{4}^{2}{{\xi }^{2}}{{C}^{2}}}{{{{\tilde{\eta }}}_{4}}{{\zeta }^{1/2}}}{{Y}_{0}}-\frac{36{{{\tilde{\eta }}}_{4}}}{{{\zeta }^{1/2}}\kappa _{4}^{2}}{{Y}_{0}}~. 	
\end{eqnarray}
Thus,  $Y_1(x)$ decreases exponentially.

\subsubsection{Asymptotic Solution in Higher Dimensions}
In the $D>4$ case, the situation is more complicated. We have to take the condition: $\lambda_1=\lambda_2=\lambda_3=0$ and
\begin{equation}\label{simplyconD}
	e^{(D-4)x} \gg   \kappa_D^2 \xi C ~ ,
\end{equation}
for $ \kappa_D^2, \xi, C$ fixed. So, at the lowest order, \eqref{mastereq} simplifies to
{\small{\begin{eqnarray}
		&&2\left( {{D}^{3}}-13{{D}^{2}}+53D-77 \right)\frac{{{Y}_{0}}^{2}}{{{\zeta }^{1/2}}\kappa _{D}^{2}}+\left\{ \frac{8(D-4)}{{{\zeta }^{1/2}}}-16\kappa _{D}^{2}{{\xi }^{2}}{{C}^{2}}{{\left( D-4 \right)}^{2}} \right\}{{Y}_{0}}+\frac{2\kappa _{D}^{2}}{{{\zeta }^{1/2}}(D-2)}=0 ~ , \notag\\
\label{eq:mastereqasymp0}
\end{eqnarray}}}
while the first order of \eqref{mastereq} has the form
{\small{\begin{eqnarray}
		&& \frac{d^2 Y_1}{dx^2} + \frac{d Y_1}{dx} {{e}^{\left( D-4 \right)x}}\left\{ 8\kappa _{D}^{2}{{\xi }^{2}}{{C}^{2}}(D-4)-\frac{4(D-2)(D-4)}{\kappa _{D}^{2}}+2-\left( \frac{2(D-2)D-4)}{\kappa _{D}^{2}}+\frac{1}{2\kappa _{D}^{2}C\xi {{\zeta }^{1/2}}} \right){{Y}_{0}} \right\} \notag\\ 
				&&+ \left\{ \frac{4\left( {{D}^{3}}-13{{D}^{2}}+53D-77 \right)}{{{\zeta }^{1/2}}\kappa _{D}^{2}}{{Y}_{0}}+\frac{8(D-4)}{{{\zeta }^{1/2}}}-16\kappa _{D}^{2}{{\xi }^{2}}{{C}^{2}}{{\left( D-4 \right)}^{2}} \right\}{{e}^{\left( D-4 \right)x}}{{Y}_{1}}=0 ~. \label{eq:mastereqasymp1}	
\end{eqnarray}}}
At the lowest order, the spacetime becomes space of constant Ricci scalar but not Einstein. Here,  we  consider some simple solutions of \eqref{eq:mastereqasymp0} and \eqref{eq:mastereqasymp1} where the constant $\zeta$ takes a particular form such that it is easy to see $Y_0 > 0$.

Let us first focus on $D=5,6$ case.  Taking  
\begin{equation}\label{zetacondition}
	\zeta =\frac{144}{{{(D-7)}^{2}}}{{Y}_{0}}^{2}~,
\end{equation}
we obtain
\begin{eqnarray}\label{Y0d56}
			{{Y}_{0}}&=&-\frac{2(D-4)(D-7)\kappa _{D}^{2}}{\left( {{D}^{3}}-13{{D}^{2}}+53D-77 \right)(D-7)-96\kappa _{D}^{4}{{\xi }^{2}}{{C}^{2}}{{\left( D-4 \right)}^{2}}} \notag\\
			\notag\\ 
		&+&\frac{\kappa _{D}^{2}\sqrt{12(D-1){{(D-7)}^{2}}\left( {{D}^{2}}-8D+17 \right)+4(D-7)96\kappa _{D}^{4}{{\xi }^{2}}{{C}^{2}}{{\left( D-4 \right)}^{2}}}}{2\sqrt{D-2}\left\{ \left( {{D}^{3}}-13{{D}^{2}}+53D-77 \right)(D-7)-96\kappa _{D}^{4}{{\xi }^{2}}{{C}^{2}}{{\left( D-4 \right)}^{2}} \right\}}~,\notag\\
		&&
\end{eqnarray}
with the constant $C$ satisfies
\begin{eqnarray}\label{C56}
	C>\frac{2\left( D-4 \right)(D-7)}{\left( {{D}^{4}}-212{{D}^{3}}+2064{{D}^{2}}-6110D+4755 \right)\xi -192\xi {{\left( D-4 \right)}^{2}}(D-2){{Y}_{0}}}
\end{eqnarray}
Then, we have the first order solution 
\begin{eqnarray}\label{Y1d56}
			{{Y}_{1}}\left( x \right) &=& c {{\left( \frac{D-4}{2\pi \sqrt{\beta {{e}^{\left( D-4 \right)x}}}} \right)}^{1/2}}\left\{ \left( 1-\frac{D-4}{16\sqrt{\beta {{e}^{\left( D-4 \right)x}}}}   \right)\cos \left( \frac{2\sqrt{\beta {{e}^{\left( D-4 \right)x}}}}{D-4} \right) \right. \notag\\ 
			&& \left. +\left( 1+\frac{D-4}{16\sqrt{\beta {{e}^{\left( D-4 \right)x}}}}  \right)\sin \left( \frac{2\sqrt{\beta {{e}^{\left( D-4 \right)x}}}}{D-4} \right) \right\} ~. 	
\end{eqnarray}
where $c$ is a real constant and 
\begin{equation}
	\beta =\frac{2(D-4)(D-7)}{3{{Y}_{0}}}+\frac{(D-2)(D-4){{(D-7)}^{2}}+3{{(D-7)}^{2}}}{3\kappa _{D}^{2}}-16\kappa _{D}^{2}{{\xi }^{2}}{{C}^{2}}{{\left( D-4 \right)}^{2}}.
\end{equation}

For $D=7$ case,  if we take the same condition as \eqref{zetacondition}, then $\zeta\to\infty$. Since $C>0$, one would have $Y_0=0$ from \eqref{eq:mastereqasymp0} implying that there is no asymptotic solution for this case.


Finally, for $D>7$ case,  taking $\zeta$ as in \eqref{zetacondition}, we find
\begin{eqnarray}\label{Y0dlebih7asimtotik}
	{{Y}_{0}}&=&-\frac{2(D-4)(D-7)\kappa _{D}^{2}}{\left( {{D}^{3}}-13{{D}^{2}}+53D-77 \right)(D-7)-96\kappa _{D}^{4}{{\xi }^{2}}{{C}^{2}}{{\left( D-4 \right)}^{2}}} \notag\\
	\notag\\ 
	&-&\frac{\kappa _{D}^{2}\sqrt{12(D-1){{(D-7)}^{2}}\left( {{D}^{2}}-8D+17 \right)+4(D-7)96\kappa _{D}^{4}{{\xi }^{2}}{{C}^{2}}{{\left( D-4 \right)}^{2}}}}{2\sqrt{D-2}\left\{ \left( {{D}^{3}}-13{{D}^{2}}+53D-77 \right)(D-7)-96\kappa _{D}^{4}{{\xi }^{2}}{{C}^{2}}{{\left( D-4 \right)}^{2}} \right\}}~,\notag\\
	&& ~ .
\end{eqnarray}
Similar to the preceding discussion, the constant $C$ fulfills
\begin{eqnarray}\label{Clebih7}
			C<\frac{2\left( D-4 \right)(D-7)}{\left( {{D}^{4}}-212{{D}^{3}}+2064{{D}^{2}}-6110D+4755 \right)\xi -192\xi {{\left( D-4 \right)}^{2}}(D-2){{Y}_{0}}}
\end{eqnarray}
Then, we get the first-order solution as in \eqref{Y1d56} for $D>7$.


Next, from \eqref{Y55} we also can get the first order scalar field 
\begin{eqnarray}
	\phi (R) &=& \phi_0 + \int \frac{dx}{\sqrt{Y_0+Y_1}} \approx \frac{1}{\sqrt{Y_0}}\int \left(1-\frac{Y_1}{2Y_0}\right)dx\notag\\
	&\approx& \phi_0 + Y_0^{-1/2} \ln R - \frac{1}{2} Y_0^{-3/2} \int Y_1 ~ dx  ~ .
\end{eqnarray}
with $Y_0$ satisfy (\ref{Y0d56}) for $D=5,6$ and \eqref{Y0dlebih7asimtotik} for $D>7$.

The first order potential scalar for $D=5,6$ and $D>7$ can be obtained from \eqref{toreq8} whose form is
\begin{eqnarray}
			V\left( R \right)&=&\frac{(D-2)\left( D-4 \right)C\left( {{R}^{\left( D-4 \right)}}+2\kappa _{D}^{2}\xi C \right)}{64\pi \beta \kappa _{D}^{2}{{R}^{3(D-4)}}}\left\{ \left( 16\beta {{R}^{\left( D-4 \right)}}-7(D-4)\sqrt{\beta }{{R}^{\frac{\left( D-4 \right)}{2}}}+{{\left( D-4 \right)}^{2}} \right) \right. \notag\\ 
			&\times& \cos \left( \frac{2\sqrt{\beta }}{\left( D-4 \right)}{{R}^{\frac{\left( D-4 \right)}{2}}} \right)-\left( 16\beta {{R}^{\left( D-4 \right)}}+7(D-4)\sqrt{\beta }{{R}^{\frac{\left( D-4 \right)}{2}}}+{{\left( D-4 \right)}^{2}} \right) \notag\\ 
			&\times& \left. \sin \left( \frac{2\sqrt{\beta }}{\left( D-4 \right)}{{R}^{\frac{\left( D-4 \right)}{2}}} \right) \right\}-\left\{ \frac{(D-7)(D-2)C}{2\kappa _{D}^{2}{{R}^{\left( D-4 \right)}}}+\frac{2\xi {{C}^{2}}(D-3)(D-2)}{{{R}^{2\left( D-4 \right)}}} \right\}{{Y}_{0}} \notag\\ 
			&-&\frac{1}{4\kappa _{D}^{2}\xi }-\frac{C}{2{{R}^{(D-4)}}}-c{{\left( \frac{D-4}{2\pi \sqrt{\beta {{R}^{(D-4)}}}} \right)}^{1/2}} \left[ \frac{(D-7)(D-2)C}{2\kappa _{D}^{2}{{R}^{\left( D-4 \right)}}}+\frac{2\xi {{C}^{2}}(D-3)(D-2)}{{{R}^{2\left( D-4 \right)}}} \right]\notag\\ 
			&\times& \left\{ \left( 1-\frac{D-4}{16\sqrt{\beta {{R}^{(D-4)}}}} \right)\cos \left( \frac{2\sqrt{\sigma {{R}^{(D-4)}}}}{D-4} \right)+\left( 1+\frac{D-4}{16\sqrt{\beta {{R}^{(D-4)}}}} \right)\sin \left( \frac{2\sqrt{\beta {{R}^{(D-4)}}}}{D-4} \right) \right\} \notag\\ ~ ,
\end{eqnarray}

 \section{Local-Global Existences of The Master Equation}
  \label{sec:localglobalex}
In this section we prove local existence and uniqueness of the maximal solution of the master equation \eqref{mastereq} using Picard's iteration and the contraction mapping theorem. Then, we prove the global existence and give some necessary conditions in order to have a regular solution on any proper interval of $\lR$.
 \subsection{Local Existence and Uniqueness }
 Let us first introduce the dynamical quantity
 \begin{equation}
 {\bf{u}}   \equiv  \left( \begin{array}{c}
Y  \\
P_Y 
\end{array} \right) ~ , \label{dynvar}
\end{equation}
where $P_Y  \equiv dY/dx$  defined on  an interval $I \equiv [x_0, x_0 + \varepsilon]$ where $x_0 \in \lR$ and $\varepsilon$ is a small positive constant. On $I$, we may have  $Y > 0$ since there is no minimal submanifold according Theorem \ref{thmmainresnohor}. Then, we can write the master equation \eqref{mastereq} as 
 \begin{equation}
 \frac{d {\bf{u}} }{dx}  = \mathcal{J}({\bf{u}}, x)   \equiv  \left( \begin{array}{c}
P_Y  \\
J_Y
\end{array} \right) ~ , \label{fungsiJ}
\end{equation}
 with
 {\footnotesize{
	 \begin{eqnarray}\label{JY}
		 				 {{J}_{Y}}&\equiv&{{P}_{Y}}^{2}\left\{ \frac{1}{2}-\frac{\kappa _{D}^{2}}{8{{\zeta }^{1/2}}(D-2)\tilde{\eta }} \right\}+{{Y}^{2}}\left\{ -\frac{2(D-2)}{{{\zeta }^{1/2}}\kappa _{D}^{2}}\tilde{\eta }{{\left( \frac{(D-7)\kappa _{D}^{2}\xi C}{(D-2)\tilde{\eta }}+\frac{2{{\left( \kappa _{D}^{2}\xi C \right)}^{2}}\left( D-4 \right)}{(D-2)\tilde{\eta }\eta }-(D-4) \right)}^{2}} \right. \notag\\ 
		 				&+&\frac{4\left( D-4 \right)\xi C}{{{\zeta }^{1/2}}\eta }\left( (D-7)\frac{\kappa _{D}^{2}\xi C}{{\tilde{\eta }}}+\frac{2{{\left( \kappa _{D}^{2}\xi C \right)}^{2}}\left( D-4 \right)}{\tilde{\eta }\eta }-\left( D-2 \right)(D-4) \right)+\frac{6(D-2)\left( D-4 \right)}{{{\zeta }^{1/2}}\kappa _{D}^{2}}\tilde{\eta } \notag\\ 
		 				&-&\left. \frac{6\tilde{\eta }}{{{\zeta }^{1/2}}\kappa _{D}^{2}}\left( (D-7)+\frac{2\kappa _{D}^{2}\xi C\left( D-4 \right)}{\eta } \right) \right\}+Y{{P}_{Y}}\left\{ 2\xi C(D-7)+\frac{4\kappa _{D}^{2}{{\xi }^{2}}{{C}^{2}}\left( D-4 \right)}{\eta }+\frac{2(D-2)D-4)\tilde{\eta }}{\kappa _{D}^{2}} \right. \notag\\ 
		 				&-&\frac{(D-7)\kappa _{D}^{2}\xi C}{(D-2){{\zeta }^{1/2}}\tilde{\eta }}-\frac{2{{\left( \kappa _{D}^{2}\xi C \right)}^{2}}\left( D-4 \right)}{(D-2){{\zeta }^{1/2}}\tilde{\eta }\eta }+\left. (D-4)+\frac{\left( D-4 \right)\kappa _{D}^{2}\xi C}{{{\zeta }^{1/2}}\eta \tilde{\eta }}+\frac{{\tilde{\eta }}}{2\kappa _{D}^{2}C\xi {{\zeta }^{1/2}}} \right\} \notag\\ 
		 				&-&{{P}_{Y}}\left\{ \left( \frac{\left( D-4 \right)}{{\tilde{\eta }}}-\frac{4\kappa _{D}^{2}\xi C}{\eta } \right){{e}^{\left( D-4 \right)x}}+4\xi C(D-7)+\left( 8\kappa _{D}^{2}{{\xi }^{2}}{{C}^{2}}-\frac{4(D-2)}{\kappa _{D}^{2}} \right)(D-4)\tilde{\eta } \right. \notag\\ 
		 				&+&\frac{2C\kappa _{D}^{4}\xi }{(D-2)}\frac{{{e}^{\left( D-4 \right)x}}}{{{\zeta }^{1/2}}\tilde{\eta }\eta }+\frac{2\kappa _{D}^{2}\xi {{\lambda }_{1}}\hat{\eta }+4\kappa _{D}^{4}{{\xi }^{2}}C{{\lambda }_{2}}}{(D-2)}\frac{{{e}^{\left( D-6 \right)x}}}{{{\zeta }^{1/2}}\tilde{\eta }\eta }-\frac{4\xi {{\lambda }_{1}}\hat{\eta }+8\kappa _{D}^{2}{{\xi }^{2}}C{{\lambda }_{2}}}{\eta }{{e}^{\left( D-6 \right)x}} \notag\\ 
		 				&+&\left( \frac{2}{\eta }+\frac{2\kappa _{D}^{2}\xi }{\eta }\left( \frac{{{\lambda }_{1}}}{2}+{{\lambda }_{2}} \right) \right){{e}^{\left( 2D-8 \right)x}}-\left. \left( \left( \frac{{{\lambda }_{1}}}{2}+{{\lambda }_{2}} \right)\frac{\kappa _{D}^{4}\xi }{(D-2)}+\frac{\kappa _{D}^{2}}{(D-2)} \right)\frac{{{e}^{\left( 2D-8 \right)x}}}{{{\zeta }^{1/2}}\tilde{\eta }\eta } \right\} \notag\\ 
		 				&+&Y\left\{ \frac{8\kappa _{D}^{2}{{\xi }^{2}}{{C}^{2}}{{\left( D-4 \right)}^{2}}\left( \tilde{\eta }+\eta  \right)}{\eta }{{e}^{\left( D-4 \right)x}}-\frac{8C\kappa _{D}^{4}\xi }{{{\zeta }^{1/2}}}\frac{{{e}^{\left( D-4 \right)x}}}{\eta } \right.+\frac{\left( D-4 \right)(D-7)\xi C}{{\tilde{\eta }}}{{e}^{\left( D-4 \right)x}} \notag\\ 
		 				&+&\left( \left( \frac{{{\lambda }_{1}}}{2}+{{\lambda }_{2}} \right)\frac{\kappa _{D}^{4}\xi }{(D-2)}+\frac{\kappa _{D}^{2}}{(D-2)} \right)\frac{4(D-2)}{{{\zeta }^{1/2}}\kappa _{D}^{2}}\frac{{{e}^{\left( 2D-8 \right)x}}}{\eta }+\frac{\left( 8\kappa _{D}^{2}\xi {{\lambda }_{1}}\hat{\eta }+16\kappa _{D}^{4}{{\xi }^{2}}C{{\lambda }_{2}} \right)}{{{\zeta }^{1/2}}}\frac{\left( D-4 \right){{e}^{\left( D-6 \right)x}}}{\tilde{\eta }{{\eta }^{2}}} \notag\\ 
		 				&-&\frac{8\xi {{\lambda }_{1}}\hat{\eta }+16\kappa _{D}^{2}{{\xi }^{2}}C{{\lambda }_{2}}}{{{\zeta }^{1/2}}}\frac{{{e}^{\left( D-6 \right)x}}}{\eta }\left( \frac{(D-7)\kappa _{D}^{2}\xi C}{(D-2)\tilde{\eta }}+\frac{2{{\left( \kappa _{D}^{2}\xi C \right)}^{2}}\left( D-4 \right)}{(D-2)\tilde{\eta }\eta }-(D-4) \right) \notag\\ 
		 				&-&\left( \left( \frac{{{\lambda }_{1}}}{2}+{{\lambda }_{2}} \right)\kappa _{D}^{4}{{\xi }^{2}}C+\kappa _{D}^{2}\xi C \right)\frac{4\left( D-4 \right){{e}^{\left( 2D-8 \right)x}}}{{{\zeta }^{1/2}}\tilde{\eta }{{\eta }^{2}}}+\frac{8\left( D-4 \right){{\left( \kappa _{D}^{2}\xi C \right)}^{2}}}{{{\zeta }^{1/2}}}\frac{{{e}^{\left( D-4 \right)x}}}{\tilde{\eta }{{\eta }^{2}}} \notag\\ 
		 				&+&\frac{6\tilde{\eta }}{\eta {{\zeta }^{1/2}}C}\left( \frac{{{\lambda }_{1}}}{2}+{{\lambda }_{2}} \right){{e}^{(2D-8)x}}-\left. \frac{8\xi {{e}^{(D-6)x}}\tilde{\eta }}{\eta {{\zeta }^{1/2}}}\left( -\frac{5{{\lambda }_{1}}}{2}+4{{\lambda }_{3}}-2(D-7){{\lambda }_{2}} \right) \right\} \notag\\ 
		 				&+&\frac{8\kappa _{D}^{2}\xi C\left( \tilde{\eta }+\eta  \right)\left( D-4 \right)}{\tilde{\eta }{{\eta }^{2}}}{{e}^{\left( 2D-8 \right)x}}-\frac{8{{\lambda }_{1}}\xi \left( \tilde{\eta }\eta -\eta \hat{\eta }-\tilde{\eta }\hat{\eta } \right)-8\kappa _{D}^{2}{{\xi }^{2}}C{{\lambda }_{2}}\left( \eta +\tilde{\eta } \right)}{\tilde{\eta }{{\eta }^{2}}}\left( D-4 \right){{e}^{2\left( D-5 \right)x}} \notag\\ 
		 				&-&\frac{8\kappa _{D}^{2}\xi C\left( D-4 \right)}{\eta }{{e}^{\left( D-4 \right)x}}+\left( \kappa _{D}^{2}+\kappa _{D}^{4}\xi \left( \frac{{{\lambda }_{1}}}{2}+{{\lambda }_{2}} \right) \right)\left( \frac{8\left( D-4 \right){{e}^{\left( 2D-8 \right)x}}}{\kappa _{D}^{2}\eta }-\frac{4\left( \tilde{\eta }+\eta  \right)\left( D-4 \right){{e}^{3\left( D-4 \right)x}}}{\kappa _{D}^{2}\tilde{\eta }{{\eta }^{2}}} \right) \notag\\ 
		 				&-&\frac{2(D-2)}{{{\zeta }^{1/2}}\kappa _{D}^{2}}\tilde{\eta }\left( \frac{2\kappa _{D}^{2}\xi {{\lambda }_{1}}\hat{\eta }+4\kappa _{D}^{4}{{\xi }^{2}}C{{\lambda }_{2}}}{(D-2)}\frac{{{e}^{\left( D-6 \right)x}}}{\tilde{\eta }\eta }-\left( \left( \frac{{{\lambda }_{1}}}{2}+{{\lambda }_{2}} \right)\frac{\kappa _{D}^{4}\xi }{(D-2)}+\frac{\kappa _{D}^{2}}{(D-2)} \right)\frac{{{e}^{\left( 2D-8 \right)x}}}{\tilde{\eta }\eta } \right. \notag\\ 
		 				&+&{{\left. \frac{2C\kappa _{D}^{4}\xi }{(D-2)}\frac{{{e}^{\left( D-4 \right)x}}}{\tilde{\eta }\eta } \right)}^{2}}-\frac{8{{\lambda }_{1}}\xi \hat{\eta }+16\kappa _{D}^{2}{{\xi }^{2}}C{{\lambda }_{2}}}{\eta }\left( D-6 \right){{e}^{\left( D-6 \right)x}}~.  			
	\end{eqnarray} } }
%
\begin{lemma}
	\label{opJY}
Let  $U \subset \lR^2$ be an open set. The nonlinear operator $ \mathcal{J}({\bf{u}}, x) $ in \eqref{JY} is locally Lipschitz with respect to $\bf{u}$ on $U$.
\end{lemma}
\begin{proof}
We have the following estimate
 {\footnotesize{
		\begin{eqnarray}\label{estJY}
			 			{{\left| {{J}_{Y}} \right|}_{U}}&\le& {{\left| {{P}_{Y}} \right|}^{2}}\left\{ \frac{1}{2}+\frac{\kappa _{D}^{2}}{8{{\zeta }^{1/2}}(D-2)\tilde{\eta }} \right\}+{{\left| Y \right|}^{2}}\left\{ \frac{2(D-2)}{{{\zeta }^{1/2}}\kappa _{D}^{2}}\tilde{\eta }{{\left( \frac{(D-7)\kappa _{D}^{2}\xi C}{(D-2)\tilde{\eta }}+\frac{2{{\left( \kappa _{D}^{2}\xi C \right)}^{2}}\left( D-4 \right)}{(D-2)\tilde{\eta }\eta }+(D-4) \right)}^{2}} \right. \notag\\ 
			 			&+&\frac{4\left( D-4 \right)\xi C}{{{\zeta }^{1/2}}\eta }\left( (D-7)\frac{\kappa _{D}^{2}\xi C}{{\tilde{\eta }}}+\frac{2{{\left( \kappa _{D}^{2}\xi C \right)}^{2}}\left( D-4 \right)}{\tilde{\eta }\eta }+\left( D-2 \right)(D-4) \right)+\frac{6(D-2)\left( D-4 \right)}{{{\zeta }^{1/2}}\kappa _{D}^{2}}\tilde{\eta } \notag\\ 
			 			&+&\left. \frac{6\tilde{\eta }}{{{\zeta }^{1/2}}\kappa _{D}^{2}}\left( (D-7)+\frac{2\kappa _{D}^{2}\xi C\left( D-4 \right)}{\eta } \right) \right\}+\left| Y{{P}_{Y}} \right|\left\{ 2\xi C(D-7)+\frac{4\kappa _{D}^{2}{{\xi }^{2}}{{C}^{2}}\left( D-4 \right)}{\eta }+\frac{2(D-2)D-4)\tilde{\eta }}{\kappa _{D}^{2}} \right. \notag\\ 
			 			&+&\frac{(D-7)\kappa _{D}^{2}\xi C}{(D-2){{\zeta }^{1/2}}\tilde{\eta }}+\frac{2{{\left( \kappa _{D}^{2}\xi C \right)}^{2}}\left( D-4 \right)}{(D-2){{\zeta }^{1/2}}\tilde{\eta }\eta }+\left. (D-4)+\frac{\left( D-4 \right)\kappa _{D}^{2}\xi C}{{{\zeta }^{1/2}}\eta \tilde{\eta }}+\frac{{\tilde{\eta }}}{2\kappa _{D}^{2}C\xi {{\zeta }^{1/2}}} \right\} \notag\\ 
			 			&+&\left| {{P}_{Y}} \right|\left\{ \left( \frac{\left( D-4 \right)}{{\tilde{\eta }}}+\frac{4\kappa _{D}^{2}\xi C}{\eta } \right){{e}^{\left( D-4 \right)x}}+4\xi C(D-7)+\left( 8\kappa _{D}^{2}{{\xi }^{2}}{{C}^{2}}+\frac{4(D-2)}{\kappa _{D}^{2}} \right)(D-4)\tilde{\eta } \right. \notag\\ 
			 			&+&\frac{2C\kappa _{D}^{4}\xi }{(D-2)}\frac{{{e}^{\left( D-4 \right)x}}}{{{\zeta }^{1/2}}\tilde{\eta }\eta }+\frac{2\kappa _{D}^{2}\xi {{\lambda }_{1}}\hat{\eta }+4\kappa _{D}^{4}{{\xi }^{2}}C{{\lambda }_{2}}}{(D-2)}\frac{{{e}^{\left( D-6 \right)x}}}{{{\zeta }^{1/2}}\tilde{\eta }\eta }+\frac{4\xi {{\lambda }_{1}}\hat{\eta }+8\kappa _{D}^{2}{{\xi }^{2}}C{{\lambda }_{2}}}{\eta }{{e}^{\left( D-6 \right)x}} \notag\\ 
			 			&+&\left( \frac{2}{\eta }+\frac{2\kappa _{D}^{2}\xi }{\eta }\left| \frac{{{\lambda }_{1}}}{2}+{{\lambda }_{2}} \right| \right){{e}^{\left( 2D-8 \right)x}}+\left. \left( \left| \frac{{{\lambda }_{1}}}{2}+{{\lambda }_{2}} \right|\frac{\kappa _{D}^{4}\xi }{(D-2)}+\frac{\kappa _{D}^{2}}{(D-2)} \right)\frac{{{e}^{\left( 2D-8 \right)x}}}{{{\zeta }^{1/2}}\tilde{\eta }\eta } \right\} \notag\\ 
			 			&+&\left| Y \right|\left\{ \frac{8\kappa _{D}^{2}{{\xi }^{2}}{{C}^{2}}{{\left( D-4 \right)}^{2}}\left( \tilde{\eta }+\eta  \right)}{\eta }{{e}^{\left( D-4 \right)x}}+\frac{8C\kappa _{D}^{4}\xi }{{{\zeta }^{1/2}}}\frac{{{e}^{\left( D-4 \right)x}}}{\eta } \right.+\frac{\left( D-4 \right)(D-7)\xi C}{{\tilde{\eta }}}{{e}^{\left( D-4 \right)x}} \notag\\ 
			 			&+&\left( \left| \frac{{{\lambda }_{1}}}{2}+{{\lambda }_{2}} \right|\frac{\kappa _{D}^{4}\xi }{(D-2)}+\frac{\kappa _{D}^{2}}{(D-2)} \right)\frac{4(D-2)}{{{\zeta }^{1/2}}\kappa _{D}^{2}}\frac{{{e}^{\left( 2D-8 \right)x}}}{\eta }+\frac{\left( 8\kappa _{D}^{2}\xi {{\lambda }_{1}}\hat{\eta }+16\kappa _{D}^{4}{{\xi }^{2}}C{{\lambda }_{2}} \right)}{{{\zeta }^{1/2}}}\frac{\left( D-4 \right){{e}^{\left( D-6 \right)x}}}{\tilde{\eta }{{\eta }^{2}}} \notag\\ 
			 			&+&\frac{8\xi {{\lambda }_{1}}\hat{\eta }+16\kappa _{D}^{2}{{\xi }^{2}}C{{\lambda }_{2}}}{{{\zeta }^{1/2}}}\frac{{{e}^{\left( D-6 \right)x}}}{\eta }\left( \frac{(D-7)\kappa _{D}^{2}\xi C}{(D-2)\tilde{\eta }}+\frac{2{{\left( \kappa _{D}^{2}\xi C \right)}^{2}}\left( D-4 \right)}{(D-2)\tilde{\eta }\eta }+(D-4) \right) \notag\\ 
			 			&+&\left( \left| \frac{{{\lambda }_{1}}}{2}+{{\lambda }_{2}} \right|\kappa _{D}^{4}{{\xi }^{2}}C+\kappa _{D}^{2}\xi C \right)\frac{4\left( D-4 \right){{e}^{\left( 2D-8 \right)x}}}{{{\zeta }^{1/2}}\tilde{\eta }{{\eta }^{2}}}+\frac{8\left( D-4 \right){{\left( \kappa _{D}^{2}\xi C \right)}^{2}}}{{{\zeta }^{1/2}}}\frac{{{e}^{\left( D-4 \right)x}}}{\tilde{\eta }{{\eta }^{2}}} \notag\\ 
			 			&+&\frac{6\tilde{\eta }}{\eta {{\zeta }^{1/2}}C}\left| \frac{{{\lambda }_{1}}}{2}+{{\lambda }_{2}} \right|{{e}^{(2D-8)x}}+\left. \frac{8\xi {{e}^{(D-6)x}}\tilde{\eta }}{\eta {{\zeta }^{1/2}}}\left| -\frac{5{{\lambda }_{1}}}{2}+4{{\lambda }_{3}}-2(D-7){{\lambda }_{2}} \right| \right\} \notag\\ 
			 			&+&\frac{8\kappa _{D}^{2}\xi C\left( \tilde{\eta }+\eta  \right)\left( D-4 \right)}{\tilde{\eta }{{\eta }^{2}}}{{e}^{\left( 2D-8 \right)x}}+\frac{8\left| {{\lambda }_{1}} \right|\xi \left( \tilde{\eta }\eta +\eta \hat{\eta }+\tilde{\eta }\hat{\eta } \right)+8\kappa _{D}^{2}{{\xi }^{2}}C\left| {{\lambda }_{2}} \right|\left( \eta +\tilde{\eta } \right)}{\tilde{\eta }{{\eta }^{2}}}\left( D-4 \right){{e}^{2\left( D-5 \right)x}} \notag\\ 
			 			&+&\frac{8\kappa _{D}^{2}\xi C\left( D-4 \right)}{\eta }{{e}^{\left( D-4 \right)x}}+\left( \kappa _{D}^{2}+\kappa _{D}^{4}\xi \left| \frac{{{\lambda }_{1}}}{2}+{{\lambda }_{2}} \right| \right)\left( \frac{8\left( D-4 \right){{e}^{\left( 2D-8 \right)x}}}{\kappa _{D}^{2}\eta }+\frac{4\left( \tilde{\eta }+\eta  \right)\left( D-4 \right){{e}^{3\left( D-4 \right)x}}}{\kappa _{D}^{2}\tilde{\eta }{{\eta }^{2}}} \right) \notag\\ 
			 			&+&\frac{2(D-2)}{{{\zeta }^{1/2}}\kappa _{D}^{2}}\tilde{\eta }\left( \frac{2\kappa _{D}^{2}\xi \left| {{\lambda }_{1}} \right|\hat{\eta }+4\kappa _{D}^{4}{{\xi }^{2}}C{{\lambda }_{2}}}{(D-2)}\frac{{{e}^{\left( D-6 \right)x}}}{\tilde{\eta }\eta }+\left( \left| \frac{{{\lambda }_{1}}}{2}+{{\lambda }_{2}} \right|\frac{\kappa _{D}^{4}\xi }{(D-2)}+\frac{\kappa _{D}^{2}}{(D-2)} \right)\frac{{{e}^{\left( 2D-8 \right)x}}}{\tilde{\eta }\eta } \right. \notag\\ 
			 			&+&{{\left. \frac{2C\kappa _{D}^{4}\xi }{(D-2)}\frac{{{e}^{\left( D-4 \right)x}}}{\tilde{\eta }\eta } \right)}^{2}}+\frac{8{{\lambda }_{1}}\xi \hat{\eta }+16\kappa _{D}^{2}{{\xi }^{2}}C{{\lambda }_{2}}}{\eta }\left( D-6 \right){{e}^{\left( D-6 \right)x}}~. 		
		\end{eqnarray} } }
In general, the function $Y$ belongs at least to a class of $C^2$-real functions, thus their values are bounded on any closed interval $I$. Since $\eta(x), \hat{\eta}(x)$, and $\tilde{\eta}(x)$ are a smooth functions, then the nonlinear operator $|\mathcal{J}({\bf{u}}, x)|_U$ is bounded on U.

Moreover, for all ${\bf{u}}, \hat{\bf{u}} \in U$ we have
 {\footnotesize{
 \!\!\!\!\!\!\!\!\!\!\!\!\!\!\!\!\!\!\!\!\!\!\!\!\!\!\!\!\!\!\!\!\!\!\!\!\!\!\!\!\!
	\begin{eqnarray}\label{estJYLps}
				{{\left| \mathcal{J}\left( u,x \right)-\mathcal{J}\left( \hat{u},x \right) \right|}_{U}}&\le& \left\{ \frac{1}{2}+\frac{\kappa _{D}^{2}}{8{{\zeta }^{1/2}}(D-2)\tilde{\eta }} \right\}\left( {{P}_{Y}}+{{{\hat{P}}}_{Y}} \right)\left| {{P}_{Y}}-{{{\hat{P}}}_{Y}} \right|+\left\{ \frac{6(D-2)\left( D-4 \right)}{{{\zeta }^{1/2}}\kappa _{D}^{2}}\tilde{\eta } \right. \notag\\ 
				&+&\frac{2(D-2)}{{{\zeta }^{1/2}}\kappa _{D}^{2}}\tilde{\eta }{{\left( \frac{(D-7)\kappa _{D}^{2}\xi C}{(D-2)\tilde{\eta }}+\frac{2{{\left( \kappa _{D}^{2}\xi C \right)}^{2}}\left( D-4 \right)}{(D-2)\tilde{\eta }\eta }+(D-4) \right)}^{2}} \notag\\ 
				&+&\frac{4\left( D-4 \right)\xi C}{{{\zeta }^{1/2}}\eta }\left( (D-7)\frac{\kappa _{D}^{2}\xi C}{{\tilde{\eta }}}+\frac{2{{\left( \kappa _{D}^{2}\xi C \right)}^{2}}\left( D-4 \right)}{\tilde{\eta }\eta }+\left( D-2 \right)(D-4) \right) \notag\\ 
				&+&\left. \frac{6\tilde{\eta }}{{{\zeta }^{1/2}}\kappa _{D}^{2}}\left( (D-7)+\frac{2\kappa _{D}^{2}\xi C\left( D-4 \right)}{\eta } \right) \right\}\left( Y+\hat{Y} \right)\left| Y-\hat{Y} \right| \notag\\ 
				&+&\left\{ 2\xi C(D-7)+\frac{4\kappa _{D}^{2}{{\xi }^{2}}{{C}^{2}}\left( D-4 \right)}{\eta }+\frac{2(D-2)D-4)\tilde{\eta }}{\kappa _{D}^{2}}+\frac{(D-7)\kappa _{D}^{2}\xi C}{(D-2){{\zeta }^{1/2}}\tilde{\eta }} \right.+(D-4) \notag\\ 
				&+&\left. \frac{2{{\left( \kappa _{D}^{2}\xi C \right)}^{2}}\left( D-4 \right)}{(D-2){{\zeta }^{1/2}}\tilde{\eta }\eta }+\frac{\left( D-4 \right)\kappa _{D}^{2}\xi C}{{{\zeta }^{1/2}}\eta \tilde{\eta }}+\frac{{\tilde{\eta }}}{2\kappa _{D}^{2}C\xi {{\zeta }^{1/2}}} \right\}\left( Y\left| {{P}_{Y}}-{{{\hat{P}}}_{Y}} \right|+{{{\hat{P}}}_{Y}}\left| Y-\hat{Y} \right| \right) \notag\\ 
				&+&\left\{ \left( \frac{\left( D-4 \right)}{{\tilde{\eta }}}+\frac{4\kappa _{D}^{2}\xi C}{\eta } \right){{e}^{\left( D-4 \right)x}}+4\xi C(D-7)+\left( 8\kappa _{D}^{2}{{\xi }^{2}}{{C}^{2}}+\frac{4(D-2)}{\kappa _{D}^{2}} \right)(D-4)\tilde{\eta } \right. \notag\\ 
				&+&\frac{2C\kappa _{D}^{4}\xi }{(D-2)}\frac{{{e}^{\left( D-4 \right)x}}}{{{\zeta }^{1/2}}\tilde{\eta }\eta }+\frac{2\kappa _{D}^{2}\xi {{\lambda }_{1}}\hat{\eta }+4\kappa _{D}^{4}{{\xi }^{2}}C{{\lambda }_{2}}}{(D-2)}\frac{{{e}^{\left( D-6 \right)x}}}{{{\zeta }^{1/2}}\tilde{\eta }\eta }+\frac{4\xi {{\lambda }_{1}}\hat{\eta }+8\kappa _{D}^{2}{{\xi }^{2}}C{{\lambda }_{2}}}{\eta }{{e}^{\left( D-6 \right)x}} \notag\\ 
				&+&\left( \frac{2}{\eta }+\frac{2\kappa _{D}^{2}\xi }{\eta }\left| \frac{{{\lambda }_{1}}}{2}+{{\lambda }_{2}} \right| \right){{e}^{\left( 2D-8 \right)x}}+\left. \left( \left| \frac{{{\lambda }_{1}}}{2}+{{\lambda }_{2}} \right|\frac{\kappa _{D}^{4}\xi }{(D-2)}+\frac{\kappa _{D}^{2}}{(D-2)} \right)\frac{{{e}^{\left( 2D-8 \right)x}}}{{{\zeta }^{1/2}}\tilde{\eta }\eta } \right\}\left| {{P}_{Y}}-{{{\hat{P}}}_{Y}} \right| \notag\\ 
				&+&\left\{ \frac{8\kappa _{D}^{2}{{\xi }^{2}}{{C}^{2}}{{\left( D-4 \right)}^{2}}\left( \tilde{\eta }+\eta  \right)}{\eta }{{e}^{\left( D-4 \right)x}}+\frac{8C\kappa _{D}^{4}\xi }{{{\zeta }^{1/2}}}\frac{{{e}^{\left( D-4 \right)x}}}{\eta } \right.+\frac{\left( D-4 \right)(D-7)\xi C}{{\tilde{\eta }}}{{e}^{\left( D-4 \right)x}} \notag\\ 
				&+&\left( \left| \frac{{{\lambda }_{1}}}{2}+{{\lambda }_{2}} \right|\kappa _{D}^{2}\xi +1 \right)\frac{4{{e}^{\left( 2D-8 \right)x}}}{{{\zeta }^{1/2}}\eta }+\frac{\left( 8\kappa _{D}^{2}\xi {{\lambda }_{1}}\hat{\eta }+16\kappa _{D}^{4}{{\xi }^{2}}C{{\lambda }_{2}} \right)}{{{\zeta }^{1/2}}}\frac{\left( D-4 \right){{e}^{\left( D-6 \right)x}}}{\tilde{\eta }{{\eta }^{2}}} \notag\\ 
				&+&\frac{8\xi {{\lambda }_{1}}\hat{\eta }+16\kappa _{D}^{2}{{\xi }^{2}}C{{\lambda }_{2}}}{{{\zeta }^{1/2}}}\frac{{{e}^{\left( D-6 \right)x}}}{\eta }\left( \frac{(D-7)\kappa _{D}^{2}\xi C}{(D-2)\tilde{\eta }}+\frac{2{{\left( \kappa _{D}^{2}\xi C \right)}^{2}}\left( D-4 \right)}{(D-2)\tilde{\eta }\eta }+(D-4) \right) \notag\\ 
				&+&\left( \left| \frac{{{\lambda }_{1}}}{2}+{{\lambda }_{2}} \right|\kappa _{D}^{4}{{\xi }^{2}}C+\kappa _{D}^{2}\xi C \right)\frac{4\left( D-4 \right){{e}^{\left( 2D-8 \right)x}}}{{{\zeta }^{1/2}}\tilde{\eta }{{\eta }^{2}}}+\frac{8\left( D-4 \right){{\left( \kappa _{D}^{2}\xi C \right)}^{2}}}{{{\zeta }^{1/2}}}\frac{{{e}^{\left( D-4 \right)x}}}{\tilde{\eta }{{\eta }^{2}}} \notag\\ 
				&+&\frac{6\tilde{\eta }}{\eta {{\zeta }^{1/2}}C}\left| \frac{{{\lambda }_{1}}}{2}+{{\lambda }_{2}} \right|{{e}^{(2D-8)x}}+\left. \frac{8\xi {{e}^{(D-6)x}}\tilde{\eta }}{\eta {{\zeta }^{1/2}}}\left| -\frac{5{{\lambda }_{1}}}{2}+4{{\lambda }_{3}}-2(D-7){{\lambda }_{2}} \right| \right\}\left| Y-\hat{Y} \right|     
	\end{eqnarray} } }
After some computations, we get then
\begin{equation}
\left|  \mathcal{J}({\bf u}, x) -  \mathcal{J}( \hat{{\bf u}}, x) \right|_U  \le C_{ \mathcal{J}}(|\bf{u}|, |\hat{\bf{u}}|) | \bf{u} - \hat{\bf{u}}| ~ ,  \label{localLipshitzcon}
\end{equation}
showing that $ \mathcal{J}$ is locally Lipshitz with respect to $\bf{u}$.

\end{proof}

Next, we  cast  \eqref{fungsiJ} into the  integral form 
\begin{equation}
{\bf{u} }(x) = {\bf{u} }(x_0) + \int_{x_0}^{x}\:\mathcal{J}\left( {\bf{u} }(s), s \right)\:ds ~  . \label{IntegralEquation}
\end{equation}
Then,  a Banach space is defined as
\begin{equation}
B \equiv \{ {\bf{u} } \in C(I,U\subset\lR^2) : \: {\bf{u} }(x_0) = {\bf{u} }_{0}, \: \sup_{x \in I}| {\bf{u} }(x)|_U \leq L_0 \}  ~ ,
\end{equation}
equipped with the norm
\begin{equation}
\|{\bf{u} }\|_{B} = \sup_{x \in I}\:|\mathbf{u}(x)|_U ~ ,
\end{equation}
where $L_0 > 0$. Introducing an operator $\mathcal{K}$ 
\begin{equation}
\mathcal{K}(\mathbf{u}(x)) = \mathbf{u}_{0} + \int_{x_0}^{x}  \mathcal{J}\left(\mathbf{u}(s), s\right) ds ~ , \label{OpKdefinition}
\end{equation}
and using Lemma \ref{opJY}, we have then the following lemma \cite{Akbar:2015jya}: 
\begin{lemma}
\label{unigueness}
Let $\mathcal{K}$ be an operator  defined in  (\ref{OpKdefinition}). There exist a positive constant $C_{L_0}$ such that for 
\begin{equation}
\varepsilon   \leq \min\left(\frac{1}{C_{L_0}},\frac{1}{C_{L_0} L_0 + \|\mathcal{J}({\bf{u}_0 })\|_B}\right) ~ ,
\end{equation}
the operator $\mathcal{K}$ is a mapping from $X$ to itself and satisfies
\begin{equation}
\left\|  \mathcal{J}({\bf u}, x) -  \mathcal{J}( \hat{{\bf u}}, x) \right\|_B  \le C_{ L_{0}}\varepsilon \| {\bf{u}} - \hat{{\bf{u}}}\|_B ~ .
\end{equation}
Therefore, the operator $\mathcal{K}$ is a contraction mapping on $I = [x_0, x_0  + \varepsilon ]$.
\end{lemma}
\noindent Since $\mathcal{K}$ is a contraction mapping, then from contraction mapping theorem, there exist of a unique fixed point of (\ref{OpKdefinition}) which proves that the differential equation (\ref{fungsiJ}) has a unique local solution.

The maximal solution can be constructed as follows. Suppose ${\bf u}(x)$ is defined on the interval $[x_0, x_m)$ where $x_m$ is a positive constant. Then, by repeating the above arguments of the local existence  with the initial condition ${\bf u}(x-x_n)$ for some $x_0  < x_n < x$ and using  the uniqueness condition to glue the solutions, we obtain the maximal solution ${\bf u}_{\mathrm{max}}(x)$ on $[x_0, L]$ such that either $L=+\infty$ and the master equation admits a global solution, or $L<+\infty$ and $\lim_{x\rightarrow L}\|{\bf{u}}(x)\|=\infty$. 

\subsection{Global Existence}
In  this final subsection we are trying show  that such a regular solution of \eqref{fungsiJ} on interval  $I = \lR$ does exist. In other words, the master equation \eqref{fungsiJ} admits a global solution. 

Let us first define an interval $I^+ \equiv I^+_L  \cup I^+_A$ where $I^+_L \equiv [x_0, L]$ and $I^+_A \equiv (L, +\infty)$ where $L$ is the maximum interval. For every $\tilde{C}\in I^+_A$, we write down  \eqref{IntegralEquation}  as
\begin{equation}
{\bf{u} }(\tilde{C}) = {\bf{u} }(x_0) + \int_{x_0}^{L}\:\mathcal{J}\left( {\bf{u} }(s), s \right) ~ ds ~ +  \int_{L}^{\tilde{C}} ~ \mathcal{J}\left( {\bf{u} }(s), s \right) ~ ds . \label{IntegralEquation1}
\end{equation}
Suppose on $I^+_A$, the functions $P_Y = dY/dx \to 0$,  and $Y \to Y_0$  where $Y_0$ is a positive constant.  In the case of $D=4$, the operator $J_Y \to 0$ on  $I^+_A$ if $Y_0$ satisfies (\ref{solY0d4}) which implies  that the third term in the right hand side in  \eqref{IntegralEquation1} bounded. On the other hand, we impose that $\lambda_1=\lambda_{2}=\lambda_{3}=0$ and $\zeta$ as in \eqref{zetacondition} for $D>4$. The operator $J_Y \to 0$ on $I^+_A $ if $Y_0$ satisfies \eqref{Y0d56} for $D=5,6$ and \eqref{Y0dlebih7asimtotik} for $D>7$. Meanwhile, for $D=7$, 
\begin{eqnarray}
	{{J}_{Y}}=\left( 16\kappa _{D}^{2}{{\xi }^{2}}{{C}^{2}}{{\left( D-4 \right)}^{2}}{{Y}_{0}} \right){{e}^{\left( D-4 \right)x}}
\end{eqnarray}

Since $C>0$, then $J_Y \to 0$ only if $Y_0=0$, implying that no regular global solution exists for this case.

Then, we define another interval  $I^- \equiv I^-_L  \cup I^-_A$ where $I^-_L \equiv [-L, x_0]$ and $I^-_A \equiv ( -\infty, -L)$. In this case, we have
\begin{equation}
{\bf{u} }(-\tilde{C}) = {\bf{u} }(x_0) + \int_{-L}^{x_0}\:\mathcal{J}\left( {\bf{u} }(s), s \right) ~ ds ~ +  \int_{-\tilde{C}}^{-L} ~ \mathcal{J}\left( {\bf{u} }(s), s \right) ~ ds . \label{IntegralEquation2}
\end{equation}
Suppose also we  have $P_Y = dY/dx \to 0$,  and $Y \to Y_c$ on $I^-_A $ where $Y_c$ is a positive constant. In the case of $D=4,5,6$, if we take $x \to -\infty$, the operator $J_Y \propto e^{-4x}$ $(D=4)$, $e^{-3x}$ $(D=5)$, $J_Y \propto e^{-6x}$ $(D=6)$, and $J_Y \propto e^{-9x}$ $(D=7)$ on  $I^-_A$ which implies \eqref{IntegralEquation2} diverges. While for $D>7$ the operator $J_Y \propto e^{-3(D-4)x}$. Thus, we also do not have a regular solution of  \eqref{IntegralEquation} on $I^-$ for $D\geq4$. 

Finally, we state our results
\begin{theorem}\label{thmlocglob}
	Suppose we have the master equation \eqref{fungsiJ} on an interval $I_{x_0} \equiv (x_0, +\infty) \subset \lR$. Then, 
	\begin{itemize}
		\item[I)]  For finite $x_0 \in \lR$, then we have the following cases: 
		\begin{itemize}
			\item [a)] \eqref{fungsiJ} admits well defined solutions if it satisfies either
			\begin{itemize}
				\item[i)] \eqref{solY0d4} for $D=4$
				\item[ii)] \eqref{Y0d56} with $C$ satisfy \eqref{C56} for $D=5,6$
				\item[iii)] \eqref{Y0dlebih7asimtotik} with $C$ satisfy \eqref{Clebih7} for $D>7$
			\end{itemize}
		\item[b)] There is no regular global solution for $D=7$ when $C>0$.
		\end{itemize}

		\item[II)]  If we take $x_0 \to -\infty$, then there is no regular global solution of \eqref{fungsiJ}  for $D\geq4$.
	\end{itemize}
\end{theorem}
%
%

\section{Conclusion}
\label{sec:conclusion}

We have discussed the scalar-torsion theory with non-minimal derivative coupling in higher dimensions. In particular, we considered a class of static spacetimes where the equations of motion can be cast into a single non-linear ordinary differential equation called the master equation. Such a result follows from the fact that the $(D-2)$-dimensional submanifold ${\mathcal S}^{D-2}$ should have constant triplet structures $\left( \hat{T},  \frac{ {\mathcal{E}^{i}}_{,i}}{\sqrt{\hat{g}}}, \frac{{{\hat{e}}^{\bar{b}}}_{i}\left( {\mathcal{F}^{ik}_{\bar{b}}}\right) _{,k}}{\sqrt{\hat{g}}} \right)$. This condition restricts ${\mathcal S}^{D-2}$ and it is still unknown the classification of such submanifolds. We have some examples such as the 2-sphere $S^2$ with $(0,-1,0)$, and the flat spaces $(D-2)$-dimensional torus $T^{D-2}$ and $\lR^{D-2}$ with $(0,0,0)$. They are spaces of constant Ricci scalar curvature.

Then, we considered two simple models, namely, the $D=4$ case with particular potential\eqref{potential4d} and the $D > 4$ case with the constant $C=0$ in (\ref{constraint}). Both cases have similar behavior at the the origin where there is real singularity, while in the asymptotic region the spacetimes converge to spaces of constant Ricci scalar curvature but not Einstein.

We also have shown that in the model there exists generally at least a naked singularity at the origin on the static spacetime $ {\mathcal M}^D$ in the presence of non-minimal derivative coupling of the scalar $\phi$ with the potential turned on. Thus, there is no physical black hole where the real singularity hidden inside the horizon. In addition, it is impossible to have a smooth spacetime everywhere. In the asymptotic region where $x \to +\infty$, $Y \to Y_0$, the asymptotic geometries converge generally to spaces of constant curvature which are not Einstein.  

Since it is very difficult to have an exact solution of the master equation \eqref{mastereq}, we used a perturbative method in which the solution of \eqref{mastereq} can be expanded as $Y(x) = Y_0 + Y_1(x)$ with $|Y_1| \ll |Y_0|$ in the asymptotic region. The function $Y_1$ is the solution of the linear version of \eqref{mastereq}, that is, \eqref{eqY1d4} for $D=4$ case and \eqref{eq:mastereqasymp1} for $D >4$ case  which turns out to be an exponential decreasing function. 

We also have established  the local-global existence and the uniqueness of the master equation \eqref{mastereq} using Picard's iteration and the contraction mapping properties. We find that the regular solution of \eqref{mastereq}  on the interval interval $I_{x_0} \subset \lR$ for finite $x_0$. If we take $x_0 \to -\infty$, then we do not have global solutions for $D \ge 4$ as stated in Theorem \ref{thmlocglob}.

\section*{Acknowledgments}
We would like to thank Ainol Yaqin for useful discussion in the early stage of the work and Andy Latief for careful reading the manuscript and correcting grammar.  The work in this paper was initially  supported by Riset ITB and PDUPT Kemendikbudristek. We acknowledge the financial support from LPDP through World Class Professor Programme 2022. B.E.G.\ would like to acknowledge the support from the ICTP through the Associates Programme (2017-2022). H.S.\ is supported by Khalifa University through a Faculty Start-Up Grant (No.\ 8474000351/FSU-2021-011) and a Competitive Internal Research Awards Grant (No.\ 8474000413/CIRA-2021-065).
%

%
%

%

\end{document}